\documentclass[preprint2]{aastex}

\newcommand{\Mo} {$M_{\odot}$}
\newcommand{\Hb} {H$\beta$}
\newcommand{\Ha} {H$\alpha$}
\begin{document}

\title{Massive star formation and tidal structures in HCG~31\footnotemark{}}

\author{\'Angel R. L\'opez-S\'anchez}
\affil{Instituto de Astrof{\'\i}sica de Canarias, E-38200, La Laguna, Tenerife, Spain}
\email{angelrls@ll.iac.es}

\author{C\'esar Esteban}
\affil{Instituto de Astrof{\'\i}sica de Canarias, E-38200, La Laguna, Tenerife, Spain}
\email{cel@ll.iac.es}

\author{M\'onica Rodr\'{\i}guez}
\affil{Instituto Nacional de Astrof\'{\i}sica, \'Optica y Electr\'onica, Apdo. Postal 51 y 216, 72000 Puebla, Mexico}
\email{mrodri@inaoep.mx} 


\begin{abstract}
We present new broad-band optical and near-infrared CCD imaging together with
deep optical intermediate-resolution  spectroscopy of the Hickson Compact Group
31. We analyze the morphology and colors of the stellar populations of the 
galaxies, as well as the kinematics, physical conditions and chemical
composition of the ionized gas in order to get a  more complete view on the
origin and evolution of the system. We estimate the ages of the most recent
star formation  bursts of the system, finding an excellent consistency among
the values obtained with different indicators and starburst models. We find
that member F hosts the youngest starburst of the group, showing a
substantial population of Wolf-Rayet stars. The chemical abundances are fairly
similar in all the members of the group despite their very different absolute
magnitudes. We argue that the use of traditional metallicity-luminosity
relations based on the absolute $B$-magnitude is not appropriate for dwarf
starburst galaxies, because their luminosity is dominated by the  transient
contribution of the starburst to the blue luminosity. We think that members E
and F of the group are candidate tidal dwarf galaxies because of their high
metallicity, their kinematics, and the absence of underlying old  stellar
populations. Finally, we propose that HCG~31 is suffering several almost
simultaneous interaction processes. The most relevant of these processes are:
(a) the merging of members A and C, that would have produced two optical tidal tails; 
and (b) a fly-by encounter between G and the A+C complex, that would have produced an
\ion{H}{1} tidal tail from the stripping of the external gas of A+C,
from which members F and E have originated. 
\end{abstract}


\section{Introduction}

\footnotetext{Based on observations made with several telescopes operated on
the islands of La Palma and Tenerife by  the Isaac Newton Group of Telescopes,
Nordic Optical Telescope and Instituto de Astrof\'\i sica de Canarias in the 
Spanish observatories of Roque de Los Muchachos and Teide of the Instituto de
Astrof\'\i sica de Canarias.}

The starburst phenomenon in galaxies was first described by \citet{SS70}, who
noted that some galaxies seem to be experiencing strong episodes of star
formation. \citeauthor{SS70} estimated that the material available for the 
production of stars would be exhausted in a very short time compared to the age
of the Universe. Since this  discovery, many studies have been performed trying
to understand the processes that trigger starburst galaxies. Wolf-Rayet (WR)
galaxies are a subset of starbursts galaxies that show broad stellar
\ion{He}{2} $\lambda$ 4686  emission in  their integrated spectra due to WR
stars, whose ages are less than 6 Myr. This spectral feature indicates the
presence  of a substantial population of this sort of massive stars and offers
the opportunity to study very young bursts.  Making use of population synthesis
models it is possible to determine the age of the WR bursts and, consequently,
the  star formation and its causes can be studied. 

In dwarf galaxies, starburst phenomena cannot be understood under the
wave-density theory because of their low masses,  so other mechanisms are
needed. One of the proposed alternative mechanisms for large scale starburst
formation is  the gas compression by shocks due to the mass lost by galactic
winds, followed by the subsequent cooling of the medium. Recent  results of
this theory \citep{Hi00} predict an intermittent behavior of the bursts in
dwarf galaxies, with long inactivity periods. This theory was first
suggested by \citet{Th91}. Others authors have proposed galaxy interactions  as
the massive star formation triggering mechanism \citep{S88}, although the
interactions cannot be with neighbor  giant galaxies in most cases
\citep*{CA93,TT95}. In fact, they seem to be more frequent with low surface
brightness  galaxies \citep*{T93,T95,T96}. This led \citet{ME99} to suggest
that interactions with other dwarf objects could  be the main star
formation triggering mechanism in dwarf galaxies and, especially in WR
galaxies. 

Compact groups of galaxies are systems with a high density of galaxies and a
low dispersion velocity, constituting the  highest density cusps in the
non-clustered large-scale structure. They were defined by Hickson (1982) using
various  richness, compactness and isolation criteria. \citet{H92} and
\citet{B96} extended the compact group  definition. The resultant compact group
population represents $\sim$1-2\% of the galaxy field and is characterized by 
aggregates of 4-8 galaxies with mean projected separations of the order of a
component galaxy diameter. The small  projected separations and observed signs
of interaction imply physical densities as high as those found in the cores of
rich clusters of galaxies. In fact, interactions between the group-members are
important in some compact groups. \citet{Mo94} found that star formation rates
(SFRs) in galaxies belonging to compact groups are somewhat higher than in 
field galaxies. \citet{MOH94} reported morphological interaction signs, like
mergers and tidal  tails, in many galaxies from \citet{H82} catalogue of
compact groups. Recent studies of interactions in compact  groups can be found
in \citet*{VIP98, IPV99, IPV01, VM97, VM98, VM01, VM02, SRV01, PAM02}. The 
finding of the WR characteristics in some of the members of these
compact groups allows the study of star formation and its triggering
mechanism with detail. Perhaps, one of the best systems to perform this 
investigation is the WR galaxy NGC 1741 and the compact group where it
belongs: HCG~31.

\subsection{Brief history of the membership}

The compact group HCG~31, at a distance of 54.8 Mpc (H$_0$=75 km s$^{-1}$
Mpc$^{-1}$; \citealt{VC92}), was  identified by \citet{H82}. It is one of the
best studied compact groups because of the peculiar morphology of its  members
that includes tidal tails, a possible merger, irregular structures, and
prominent starbursts in its  brightest  galaxies (see Fig.~\ref{fig1}).
\citet{H82} described the group as four galaxies in close proximity, that he
called A,  B, C and D. He noted that two of them, members A and C, are clearly
interacting, and they both actually form NGC 1741  (Mrk 1089, VV524, Arp259).
In fact, \citet{MB93} also classified NGC 1741 as a double nucleus Markarian
galaxy.  \citet{KS86} detected the 4686 \AA\ WR feature in the spectrum of the
nucleus of galaxy C. This was later  confirmed by \citet*{R90}. NGC 1741 was
included in the first catalogue of WR galaxies by \citet{C91} as one  of the
most luminous WR galaxies known. \citet{VC92} obtained the optical spectra of A
and C and found that they are dominated by strong nebular emission lines, as
expected from the presence of a large number of hot, massive  stars. UV studies
performed by \citet*{C96} revealed that the centre of NGC 1741 is dominated by 
two main starbursts, which are composed of several intense knots of recent star
formation. \citet{J99} obtained {\it Hubble Space Telescope} ($HST$) images of
the system, dating the main starbursts (A and C) in 5 Myr,  and detecting 434
Super Star Clusters (SSCs) in the central knots. \citet{R03} called NGC 1741
the A+C complex, because they considered that A and C are a single kinematical
entity.  

Member B was classified as a Sm galaxy by \citet{H82}. It shows three different
bright knots in H$\alpha$  images (\citealt{R90, IPV97}, hereinafter IV97;
\citealt{JC00}). Its morphology suggests that it is a dwarf spiral or 
irregular galaxy seen nearly edge-on. \citet{R90} and \citet{R03} found that it
shows solid-body rotation and seems to  be kinematically distinct from the A+C
complex. 

\citet{R90} identified two nearby additional objects, E and F, and the close
galaxy Mrk 1090 (object G), that were  all included as new members of HCG~31,
as well as the far member Q, that lies 2\arcmin\ to the north of the A+C
complex.  These  authors also determined that D is a background galaxy at a
distance of 359 Mpc. IV97 studied the ages of the bursts,  and concluded that F
is as young as member C. The \ion{H}{1} map of the group \citep*{W91} shows
that  all the galaxies, except D, are embedded in the same neutral cloud.
\citeauthor{W91} estimated a total hydrogen mass of 2.1  $\times$ 10$^{10}$
M$_\odot$. Maxima of \ion{H}{1} column density are coincident with the
galaxies, indicating  that they are gas rich. 

\citet*{HCZ96} proposed several tidal dwarf candidates in HCG~31: one of them
corresponds to member F, and  the other ones are located at the northeast of the A+C
complex. \citet{JC00} considered E and F as tidal dwarf  galaxies, and detected
new SSCs in them. Those authors  confirmed that F shows a starburst of around 4
Myr, although  it can be younger. \citet{IPV01} noted 2 tidal objects at the northeast
of the A+C complex and 7 tidal objects (including E, F1, F2, and faint H) at
the southeast tidal tail, which appears delineated in the \ion{H}{1} map.
\citet{IPV01} analyzed  those objects and concluded that only member F
satisfies the escape and self-gravitation conditions to be a tidal  dwarf
galaxy. \citet{R03} argued that E is an integral part of the A+C complex or is
now separating from it. For those authors, the kinematics of member E confirms
that the extension of the A+C complex towards E and F is a tidal tail. 
\citeauthor{R03} also suggested that galaxy F could be a tidal fragment of the
A+C complex that detached some time ago.

G has a spherical shape in the broad-band images, but IV97 found that it
hosts some knots of stellar formation at  the northwest. This was confirmed by
\citet{JC00}, who found a very asymmetric H$\alpha$ emission, with the 
star-forming regions forming a U shape along the northwest side. \citet{R03} described
G as a small disk galaxy  seen nearly face-on because the H$\alpha$ velocity
has only a very small gradient across its disk. They also noted  that its
kinematics is distinct from that of the A+C complex.

\subsection{Structure of this paper}

In \S~2 we present our observations and the data reduction processes. In
\S~3 we present our optical and near-infrared photometric results,
studying the effect of the reddening in the final data.  In that section, we
also present the results of our intermediate-resolution spectroscopy: physical
conditions,  chemical abundances, and kinematics of the ionized gas. In \S~4
we discuss our results: we analyze the ages of the bursts and the stellar and WR
populations, the star formation rates, the luminosity-metallicity relation in 
the members of HCG~31, the possibility that members E and F are tidal dwarf
galaxies, and a revision of the star formation history in HCG~31. We finally
present our main conclusions in \S~5. 

\section{Observations and data reduction}


\begin{table*}[t!]\centering
  \caption{Summary of observations}
  \smallskip
  \label{table1}  
  \footnotesize
  \begin{tabular}{ccccccccc}
    \tableline\tableline
	\noalign{\smallskip}
    Observations & Telescope & Date & Exp. Time & Spatial & Filter/ & P.A. & Spectral & $\Delta\lambda$ \\
                 &           &      &    (s)  & ($\arcsec$ pix$\rm^{-1}$)& grating &($\rm^{\circ}$) & (\AA\ 
pix$\rm^{-1}$)& (\AA) \\
	\tableline
	\noalign{\smallskip}
    Broad-band & 2.56m NOT & 02/10/23 & 3 $\times$ 300 & 0.176 & $U$ & \nodata &\nodata & \nodata \\
    imaging    & 2.56m NOT & 02/10/23 & 3 $\times$ 300 & 0.176 & $B$ & \nodata &\nodata & \nodata \\
               & 2.56m NOT & 02/10/23 & 4 $\times$ 300 & 0.176 & $V$ & \nodata &\nodata & \nodata \\
		   & 2.50m INT & 03/09/22 & 2 $\times$ 200 & 0.33  & $R$ & \nodata &\nodata & \nodata \\
	       & 1.55m CST & 03/02/04 & 120 $\times$ 20& 1.0   & $J$ & \nodata &\nodata & \nodata \\
	       & 1.55m CST & 03/02/04 & 240 $\times$ 10& 1.0   & $H$ & \nodata &\nodata & \nodata \\
	       & 1.55m CST & 03/02/04 & 360 $\times$ 5 & 1.0   & $K_S$ &\nodata &\nodata & \nodata \\
    \tableline
	\noalign{\smallskip}
    Intermediate & 4.2m WHT & 00/12/29& 1800 x 4 & 0.20 & R600B&  61.0 & 0.45 & 3650-5100\\
    resolution   & 4.2m WHT & 00/12/29& 1800 x 4 & 0.36 & R136R&  61.0 & 1.49 & 5300-6650\\
    spectroscopy & 4.2m WHT & 00/12/30& 1800 x 4 & 0.20 & R600B& 128.0 & 0.45 & 3600-5200\\
                 & 4.2m WHT & 00/12/30& 1800 x 4 & 0.36 & R136R& 128.0 & 1.49 & 5500-6850\\
		 & 4.2m WHT & 00/12/31& 1800 x 4 & 0.20 & R600B& 133.0 & 0.45 & 3660-5050 \\
		 & 4.2m WHT & 00/12/31& 1800 x 4 & 0.36 & R136R& 133.0 & 1.49 & 5450-6850 \\
    \noalign{\smallskip}
	\tableline\tableline
  \end{tabular}
\end{table*}


\subsection{Optical imaging}

The images in the $U$, $B$, and $V$ filters were taken on 2002 October 23 at
the 2.56m Nordic Optical Telescope (NOT) at  Roque de los Muchachos Observatory
(La Palma, Canary Islands, Spain) during a Spanish Service Night. We used 
ALFOSC (Andalucia Faint Object Spectrograph and Camera) in image mode with a
Loral/Lesser CCD detector (2048 $\times$ 2048 pixels) with  a pixel size of 15
$\mu$m and spatial resolution of 0.188\arcsec\ pixel$^{-1}$, and the standard
Johnson filters $U$, $B$ and $V$.  Three or four 300 s exposures were added for
each filter to obtain a good signal-to-noise and an appropriate  removal of
cosmic rays in the final images. The measured full width half maximum (FWHM) of
the point-spread function (PSF) was approximately 2.2\arcsec. Images were taken
under photometric conditions, and the standard field 98-1119 of  \citet{L92}
was used to flux calibrate them. Twilight images of different zones of the sky
were taken for each filter  in order to perform the flat-field correction. The
bias subtraction, flat-fielding and flux calibration of the images  were made
following standard procedures. All the reduction process was done with
IRAF\footnote{IRAF is distributed by  NOAO which is operated by AURA Inc.,
under cooperative agreement with NSF}. We obtain the photometry of the integrated 
flux inside the 3 $\sigma$ contours of each member of HCG~31 over the sky background level.

The image in the $R$ (Sloan-Gunn) filter was taken on 2003 September 22 at the
2.50m Isaac Newton Telescope (INT) at  Roque de los Muchachos Observatory (La
Palma, Canary Islands, Spain). We used the WFC (Wide Field Camara) with a CCD 
detector (2048 $\times$ 4096 pixels) with a pixel size of 15$\mu$m and spatial
resolution of 0.33\arcsec\ pixel$^{-1}$ at  prime focus. Two 200 s exposures
were added to remove cosmic-rays. The measured FWHM of PSF was approximately 
1.1\arcsec. The standard fields 93-424 and 97-284 of \citet{L92} were used to flux
calibrate the final image. All the reduction process was done with IRAF
following the same procedure explained above.

\subsection{Near-Infrared imaging}

We used the 1.55m Carlos S\'anchez Telescope (CST) at Teide Observatory
(Tenerife, Canary Islands, Spain) to obtain  the near-infrared images on 2003
February 3. We used the CAIN camera (256 $\times$ 256 pixels) with a pixel size
of 40  $\mu$m and a spatial resolution of 1\arcsec\ pixel$^{-1}$ in the wide
mode to obtain images in the $J$ (1.2 $\mu$m), $H$ (1.6  $\mu$m) and $K_S$
(2.18$\mu$m) broad-band filters. We took 20 series of 6 consecutive individual
20 s exposures in $J$, 20 series of 12 individual 10 s exposures in $H$ and 30
series of 12 exposures of 5 s duration  in $K_S$.  Each sequence of exposures
was made at slightly different positions to obtain a clean sky image. In this
way, we  obtained a final image of 20 minutes in $J$ and $H$, and a final one
of 10 minutes in $K_S$. We repeated this  procedure two times for $J$ and $H$
and three times for $K_S$, and then combined all the images to obtain a single
image for each filter. The FWHM PSFs of these final images were 2.4\arcsec,
2.2\arcsec, and 1.9\arcsec\ for $J$, $H$, and $K_S$, respectively. Bright and 
dark dome flat-field exposures were taken for each filter, and were combined to
obtain a good flat-field image. The  standard stars As13 and As19 \citep{Hu98}
were used for the flux calibration. 

\subsection{Intermediate-resolution spectroscopy}

Intermediate-resolution spectroscopy of different galaxies of the group at
three slit positions was carried out on  2000 December 29, 30 and 31 with the
4.2m William Herschel Telescope (WHT) at Roque de los Muchachos Observatory
(La  Palma, Canary Islands, Spain) with the ISIS spectrograph at the Cassegrain
focus. Two different CCDs were used at the  blue and red arms of the
spectrograph: an EEV CCD with a configuration of 4096 $\times$ 2048 pixels of
13 $\mu$m in  the blue arm and a TEK with 1024 $\times$ 1024 of 24 $\mu$m in
the red arm. The dichroic used to separate the blue  and red beams was set at
5400 \AA. The slit was 3.7\arcmin\ long and 1\arcsec\ wide. Two gratings were
used, R600B  in the  blue arm and R316R in the red arm. These gratings
give reciprocal dispersions of 33 and 66 \AA\ mm$^{-1}$, and  effective spectral
resolutions of 2.0 and 3.9 \AA\ for the blue and red arms, respectively. The
blue spectra cover  from 3600 to 5200 \AA\ and the red ones from 5400 to 6800
\AA. The spatial resolutions were 0.20\arcsec\ pixel$^{-1}$ in the  blue and
0.36\arcsec\ pixel$^{-1}$ in the red. 

Three slit positions of HCG~31 were observed at different position angles,
which were chosen in order to cover different members of the group. For each
slit position, four 30 minutes exposures were taken and combined to obtain 
good signal-to-noise and an appropriate removal of cosmic rays in the final
blue and red spectra. Comparison lamp exposures of CuAr
for the blue arm and CuNe for the red one were taken after each set of 
spectra. The correction for atmospheric extinction was performed using an
average curve for the continuous  atmospheric extinction at Roque de los
Muchachos Observatory. The observations with slit positions at PA 61$^\circ$ 
and PA 128$^\circ$ were made at air masses very close to 1. The observations with 
slit position at PA 133$^\circ$ had air masses between 1.3 and 1.8 but were made
very close to the parallactic angle, which varied between 146$^\circ$ and 126$^\circ$
during the observation. Consequently, no correction 
was made for atmospheric differential refraction. 

The absolute flux calibration of the spectra was achieved with observations of
the standard stars G191 B2B and Feige 34 \citep{M88}. 
IRAF software was used to reduce the CCD frames (bias
correction, flat-fielding, cosmic-ray rejection, wavelength and flux
calibration, sky subtraction) and extract the one-dimensional spectra 
for each member of HCG~31. For each two-dimensional spectra several apertures were defined along the spatial direction 
to extract the final one-dimensional spectra of each galaxy or emission knot. The apertures were centered at the 
brightest point  of each  aperture and the width was fixed to obtain a good signal-to-noise spectrum. The internal 
velocity structure of the knots was not considered because it is analyzed in detail in \S 3.3.4. All apertures were 
defined on the "blue" frames. Identical apertures were then used to extract the spectra of the "red" frames. Small 
two-dimensional distortions were corrected fitting the maxima in the emission of [\ion{O}{2}]
$\lambda\lambda$3726, 3729 doublet and \Hb\ in the "blue" frames and fitting the continuum emission in the "red" 
frames.  

IRAF software was used to analyze the one-dimensional spectra (see \S3.3), but  
we also used the Starlink DIPSO  software \citep{HM90} to
analyze the profiles of selected bright emission lines as well as those zones 
where the line profiles were complex, specially the [\ion{O}{2}]
$\lambda\lambda$3726, 3729 doublet and the blend of  [\ion{Ne}{3}]
$\lambda$3967 and H$\epsilon$ lines. For each single or multiple Gaussian fit,
DIPSO gives the fit  parameters (radial velocity centroid, Gaussian sigma, FWHM,
etc) and their associated statistical errors.

The journal of all the imaging and spectroscopical observations can be found in
Table~\ref{table1}.

\section{Results}

\subsection{Optical imaging}

\begin{figure*}[!ht]
\includegraphics[width=1\linewidth]{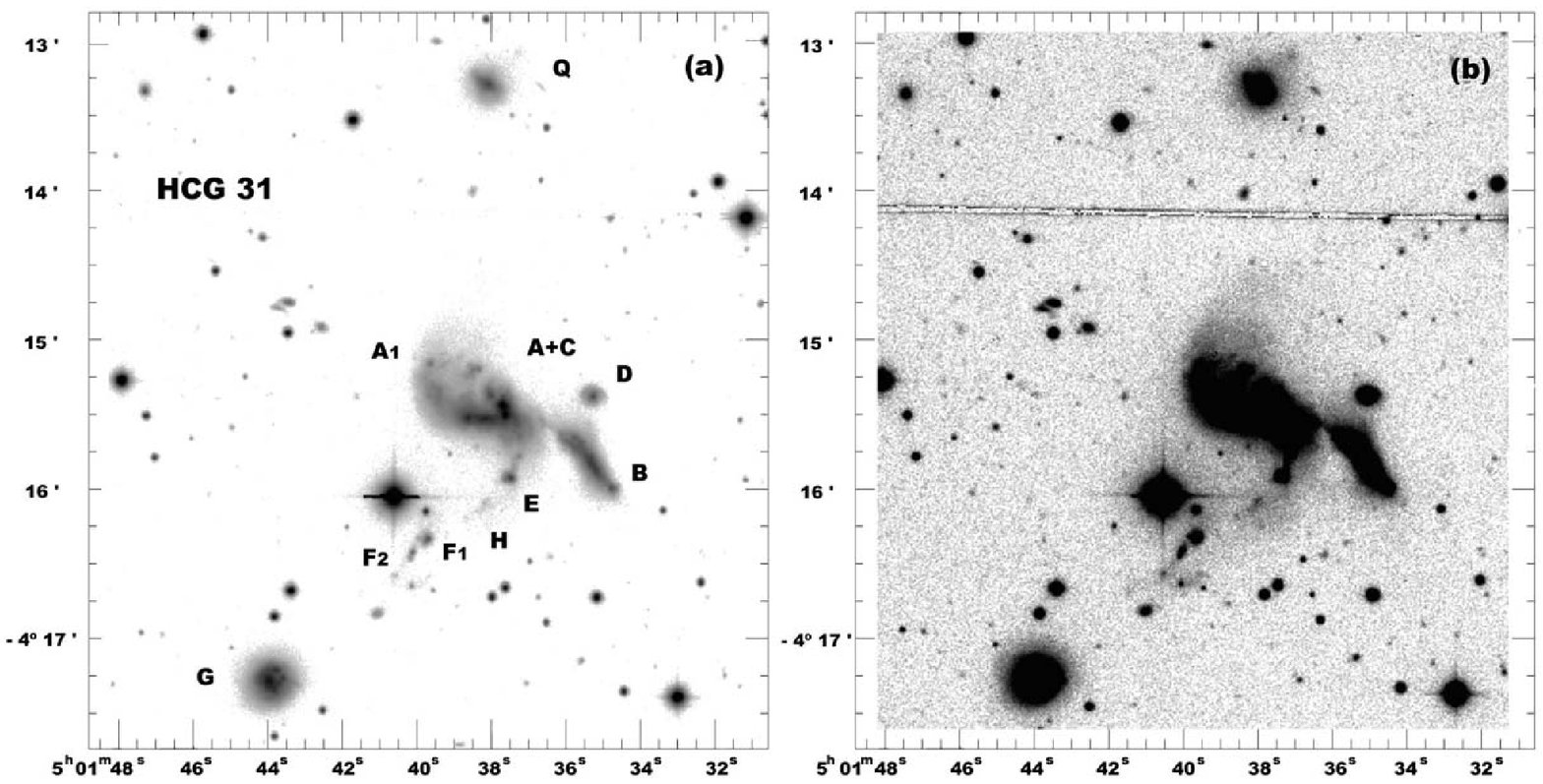}
\caption{\small{Deep optical image of HCG~31 in the $R$ filter. All the
individual galaxies of the group, as well as knot A1 (see text), are
indicated and labeled. Image (b) is a saturated version of (a) showing the
faintest  structures.}}
\label{fig1}
\end{figure*}

In Figure~\ref{fig1} we show the deep $R$ image of the system. We can
distinguish all the component members of  HCG~31, from Q that lies 2\arcmin\ to
the north of the A+C complex to member G to the south. We have labeled all the 
members of the system, as  well as some additional interesting zones that we
have also analyzed. The system is dominated by two strong bursts at  the center
of the A+C complex and a bar that extends to the east, that corresponds to the
main body of member A. Two  bright zones can be detected in the bar at our
spatial resolution. Member A also shows several additional knots  delineating a
possible tidal tail at the northeast; the zone labeled as A1 is located at the
tip of that structure (although Figure~\ref{fig1}b shows that the structure or
plume extends further to the north).  The A+C complex is clearly physically
connected with members B (at the west) and E (at the south) by faint tails or 
bridges of matter. The extension connecting the A+C complex and member  E
delineates a clearly defined arc that we call the southwest tail. Apparently,
this arc ends  at the position of member H or far away towards F. We can
appreciate  that B consists basically of three main knots that \citet{JC00}
analyzed in H$\alpha$ imaging. Member G also shows a  complex morphology with
several star-forming regions distributed at its  northwest half. Finally,
members F1 and F2 can be clearly noted. Their position near a bright star
($m_B$=11.9)  has usually been a problem for their photometry, but our new deep
images partially solve this problem. We have also  marked the position of the
faint object H that was detected in our intermediate-resolution spectra. It can
also be  noted in a previous $R$ band image obtained by \citet{IPV01}. There are
some small non-stellar faint objects surrounding  several members of the group,
specially to the northeast of the A+C complex and at the south of F1 and F2. 


The results of aperture photometry of the optical images for the different
members of the group are shown in  Table~\ref{table2}. We have also included
the values for the knot A1 because this is the zone inside member A for which
we have the spectra (see section 3.1). The data are reddening-corrected, as we
will discuss below. The area for which we have integrated the flux for each 
member is irregular and it was defined by the 3$\sigma$ level isophote in the
$B$ image. We have used the same integration  area for each object and filter
to obtain homogeneous photometric data. Galaxies A and C were measured both
together  and separately, although it is difficult to define the border area
between them. Sky subtraction was performed independently for each member
and filter. The photometry for galaxies F1 and F2 is less accurate than for the
rest of the objects due to contamination by the  bright foreground star. H is
not detected in $U$, $B$ or $V$, but we have estimated a lower limit to its 
magnitudes. In any case, we detect H in the $R$ filter (see Fig.~\ref{fig1}),
with a magnitude of 20.7$\pm$0.1.


\begin{table*}[t]
  \caption{Results of optical and near-infrared aperture photometry}
  \label{table2}  
  \scriptsize
  \begin{tabular}{llcccccccc}
    \\
    \tableline \tableline
	\noalign{\smallskip}
    Knot & $E(B-V)$   &$m_B$         &$m_J$         &$U-B$         & $B-V$       & $V-R$                             
& $V-J$ &$J-H$  &$H-K_S$ \\
    \tableline
	\noalign{\smallskip}
    A    & 0.10       &15.31$\pm$0.06&13.96$\pm$0.05&$-$0.41$\pm$0.12&   0.01$\pm$0.12 & 0.43$\pm$0.11                 
&1.34$\pm$0.12 & 0.28$\pm$0.10& 0.14$\pm$0.12 \\
    A1   & 0.10       &17.92$\pm$0.05&16.80$\pm$0.06&$-$0.58$\pm$0.10&$-$0.11$\pm$0.10 & 0.13$\pm$0.11 
&1.23$\pm$0.12 & 0.54$\pm$0.12& 0.29$\pm$0.14\\
    B    & 0.18       &14.95$\pm$0.04&14.64$\pm$0.05&$-$0.38$\pm$0.09&   0.17$\pm$0.08 & 0.06$\pm$0.08               
&0.14$\pm$0.10 & 0.13$\pm$0.10& 0.12$\pm$0.10 \\
    C    & 0.06       &14.17$\pm$0.06&14.12$\pm$0.05&$-$0.66$\pm$0.12&$-$0.01$\pm$0.12 & 0.09$\pm$0.11                   
&0.06$\pm$0.12 & 0.12$\pm$0.10& 0.21$\pm$0.12 \\
    A+C  & 0.12\tablenotemark{a}&13.59$\pm$0.04&13.36$\pm$0.05&$-$0.60$\pm$0.09&   0.03$\pm$0.08 & 0.12$\pm$0.10                   
&0.20$\pm$0.10 & 0.13$\pm$0.10& 0.15$\pm$0.10 \\
    D    & 0.15\tablenotemark{b}&18.47$\pm$0.04&16.60$\pm$0.05&   0.36$\pm$0.09&   0.68$\pm$0.08 & 0.47$\pm$0.09                   
&1.19$\pm$0.12 & 0.27$\pm$0.10& 0.53$\pm$0.10 \\
    E    & 0.06       &17.90$\pm$0.05&17.64$\pm$0.05&$-$0.65$\pm$0.10&$-$0.03$\pm$0.10 & 0.20$\pm$0.09                  
&0.29$\pm$0.12 & 0.05$\pm$0.10& 0.18$\pm$0.12 \\
    F1   & 0.20       &17.81$\pm$0.06&18.05$\pm$0.07&$-$0.99$\pm$0.12&$-$0.07$\pm$0.12 &$-$0.04$\pm$0.10                    
&$-$0.17$\pm$0.14 & 0.04$\pm$0.17& 0.29$\pm$0.30 \\
    F2   & 0.09       &19.23$\pm$0.06&19.30$\pm$0.10&$-$1.01$\pm$0.12&$-$0.09$\pm$0.12 &$-$0.02$\pm$0.10                   
&0.01$\pm$0.16 & 0.08$\pm$0.30& 0.20$\pm$0.50 \\
    G    & 0.06       &14.71$\pm$0.04&14.27$\pm$0.05&$-$0.43$\pm$0.09&$-$0.01$\pm$0.08 & 0.14$\pm$0.08                              
&0.45$\pm$0.10 & 0.12$\pm$0.10& 0.13$\pm$0.10\\
    Q    & 0.15\tablenotemark{b}&16.51$\pm$0.06&15.83$\pm$0.05&   0.07$\pm$0.12&   0.11$\pm$0.12 & 0.24$\pm$0.10                             
& 0.77$\pm$0.12 & \nodata&\nodata\\
    H\tablenotemark{c}    & 0.09&$>$20.5       & $>$19.5      &\nodata&\nodata&\nodata&\nodata&\nodata&\nodata\\
    \noalign{\smallskip}
	\tableline\tableline
	\end{tabular}
  \tablenotetext{a}{Average of the value obtained for members A, B, and C.}
  \tablenotetext{b}{Adopted from IV97.}
  \tablenotetext{c}{$m_{R}$=20.7$\pm$0.1.}
\end{table*}


We have corrected the observed photometric data for reddening. There are two
important contributions to the total  extinction by interstellar dust for an
extragalactic object: the extinction associated with the  dust of the Milky Way
in the line of sight between our Galaxy and the studied object (Galactic 
extinction) and that due to the dust mixed with the gas of the distant galaxy
(extragalactic extinction). From maps  of infrared dust emission by
\citet*{SFD98}, we know that the Galactic contribution in the  direction of
HCG~31 is $E(B-V)=0.05$, so this fixes a lower limit to the reddening
correction. A previous analysis  of the extinction toward HCG~31 was presented
by IV97, who discussed the results of three papers in which the  extinction is
analyzed: \citet{R90} and \citet{MB93}, both using the H$\alpha$/H$\beta$
ratio, and  \citet{W91}, who obtain radio \ion{H}{1} data. IV97 took these
\ion{H}{1}-based extinction values as a lower limit to the true ones, since the
\ion{H}{1} values can only account for the extinction due to dust mixed with
the neutral gas in front  of the region considered, but not within the
\ion{H}{2} region. In this way, IV97 assume $E(B-V)=0.2$ for member B  and
$E(B-V)=0.15$ for the rest of the objects in HCG~31. 

In our case, we have used the reddening constant, C(H$\beta$) -obtained from
the absorption-corrected intensities of  H$\gamma$ and H$\beta$ in our optical
spectra of each member- to correct our photometric data. The method used to 
obtain C(H$\beta$) from our spectra is described in \S~3.3. Finally, we use the
relation between C(H$\beta$) and  the extinction in $V$, $A_V$, obtained by
\citet{KL85} and assume the standard ratio of  $A_V/E(B-V)=3.1$. The data in
$B$, $U$ and $R$ were corrected using the \citet{RL85} extinction  correction
law for $A_B$, $A_U$ and $A_R$ respectively. We have adopted the reddening
values assumed by IV97 for members D and Q because we do not have spectra for
them. For the A+C complex we have assumed the average extinction  obtained for
A, B and C. The final adopted color excess, $E(B-V)$, for each individual
galaxy is shown in  Table~\ref{table2}. We have determined the errors for the
photometry of each member of HCG~31 considering the FWHM of the PSF, the sky level, and the
flux calibration error for each frame. 

\subsection{Near-Infrared imaging}

\begin{figure*}[ht]
\centering
\includegraphics[width=1\linewidth]{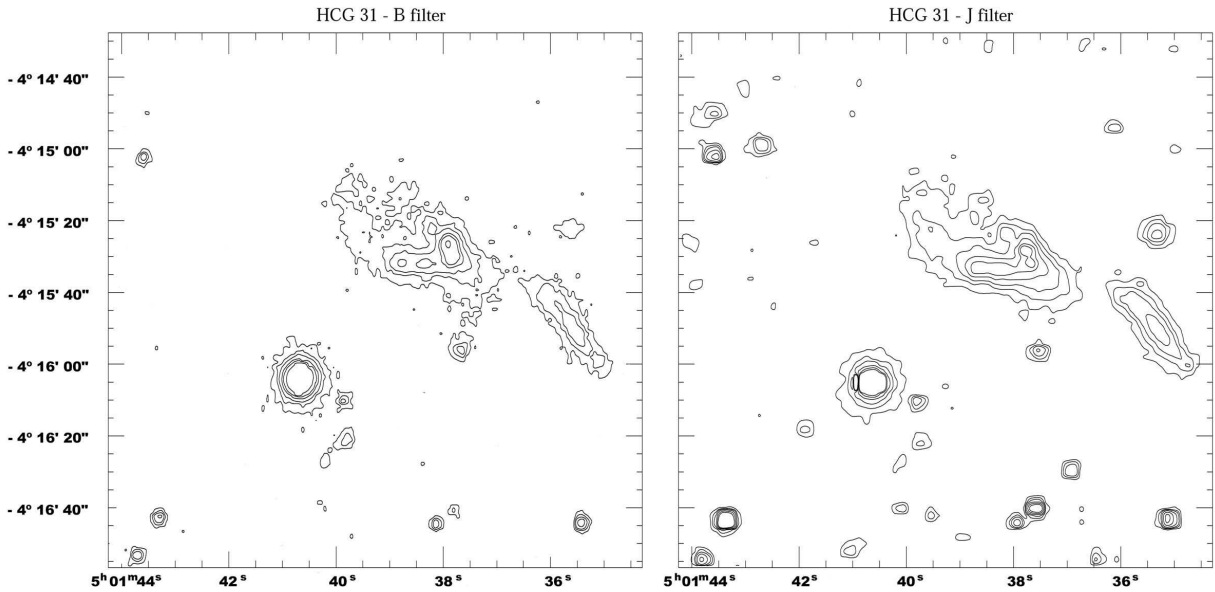}
\caption{ \small{Logarithmic contour maps of the central part of HCG~31 in $B$
and $J$ filters. The external contour corresponds to the 3$\sigma$ level in both maps.}}
\label{fig2}
\end{figure*}

Near-infrared (hereinafter NIR) photometry of HCG~31 in the $JHK$ bands was
presented in the survey of 180 interacting  galaxies carried out by
\citet{BS92}. However, they only obtained relatively low signal-to-noise data 
for members A, C, and B, so their photometric values were not very accurate. In
Figure~\ref{fig2}, we show the   logarithmic contour map of HCG~31 in the $J$
filter, and in the $B$ optical filter for comparison purposes. We have not
observed  member Q in the $H$ and $K_S$ filters. The peculiar morphology of HCG
31 is also recognized in the NIR images. The northeast  tail is evident in
$J$, but its brightness is considerably reduced in the $H$ and $K_S$ bands. In
fact, it is rather  difficult to detect A1 in $K_S$. The double nucleus and the
bar inside the A+C  complex appear bright and easily resolved in $K_S$. The
southwest tail is not detected in the NIR images, although  member E  is
clearly detected. In the NIR, the brightest zone of member B is the centre of
the galaxy, while its western knot  has practically disappeared in $K_S$.
Galaxies F1 and, especially, F2 are difficult to distinguish in $H$ and $K_S$
because of their  faintness. The morphology of G in the NIR seems very similar
to the optical morphology but the southeast half is brighter in the $K_S$ filter. We
have not detected member H but we have estimated a lower limit for its 
magnitude in each filter.   

The results of aperture photometry of the NIR images are also shown in
Table~\ref{table2}. The NIR aperture photometry  has been performed in the same
way as in the case of the optical images. In this case, the shape and size of
the apertures were  defined by the 3$\sigma$ level isophote in the $J$ image, and were
very similar to the apertures derived for the optical photometry
(see the 3$\sigma$ contours for $B$ and $J$ filters in Figure~\ref{fig2}).
We have used the same area for each NIR filter. We have also corrected the
observed photometric data for reddening following the same method used for the
optical photometry, adopting the \citet{RL85} extinction corrections for
$A_J$, $A_H$ and $A_{Ks}$ from our calculated $A_V$ value. The NIR
photometric errors were calculated following the method explained above for the
optical images. The error in $V-J$ color due to the difference in the shape and size 
between the optical and the NIR images is small when compared with the other uncertainties, 
but it has also been included.


\subsection{Intermediate resolution spectra}

\begin{figure*}[h!]
\centering
\includegraphics[width=0.8\linewidth]{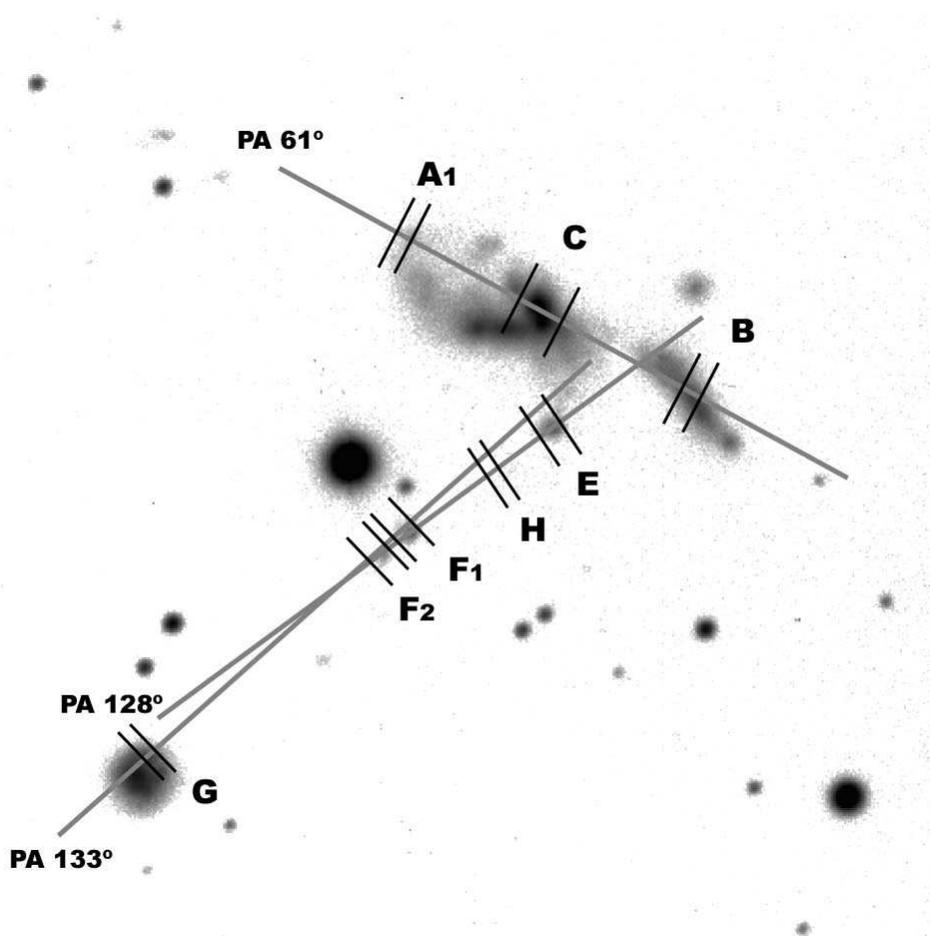}
\protect\caption[ ]{\small{Slit positions over $V$ image. The spatial extension
of the  different zones from which we have extracted the one-dimensional
spectra are indicated with ticks.}}
\label{fig3}
\end{figure*}

Figure~\ref{fig3} shows the three slit positions observed with
intermediate-resolution spectroscopy in HCG~31 over our  $V$ image. We also
indicate the different position angles (PA) observed and the location and size
of the regions  extracted in order to study the physical conditions and
chemical abundances of their ionized gas. The slit at PA  61$^\circ$ covers A1
and members C\footnote{Following the notation of previous authors, we have called C to the central zone of the A+C 
complex; it includes the two strong bursts at its center.} and B; the slit at 128$^\circ$ covers E, H and F, and the
one at 133$^\circ$  covers members F and G.

\begin{figure*}[th]
\centering
\includegraphics[width=1\linewidth]{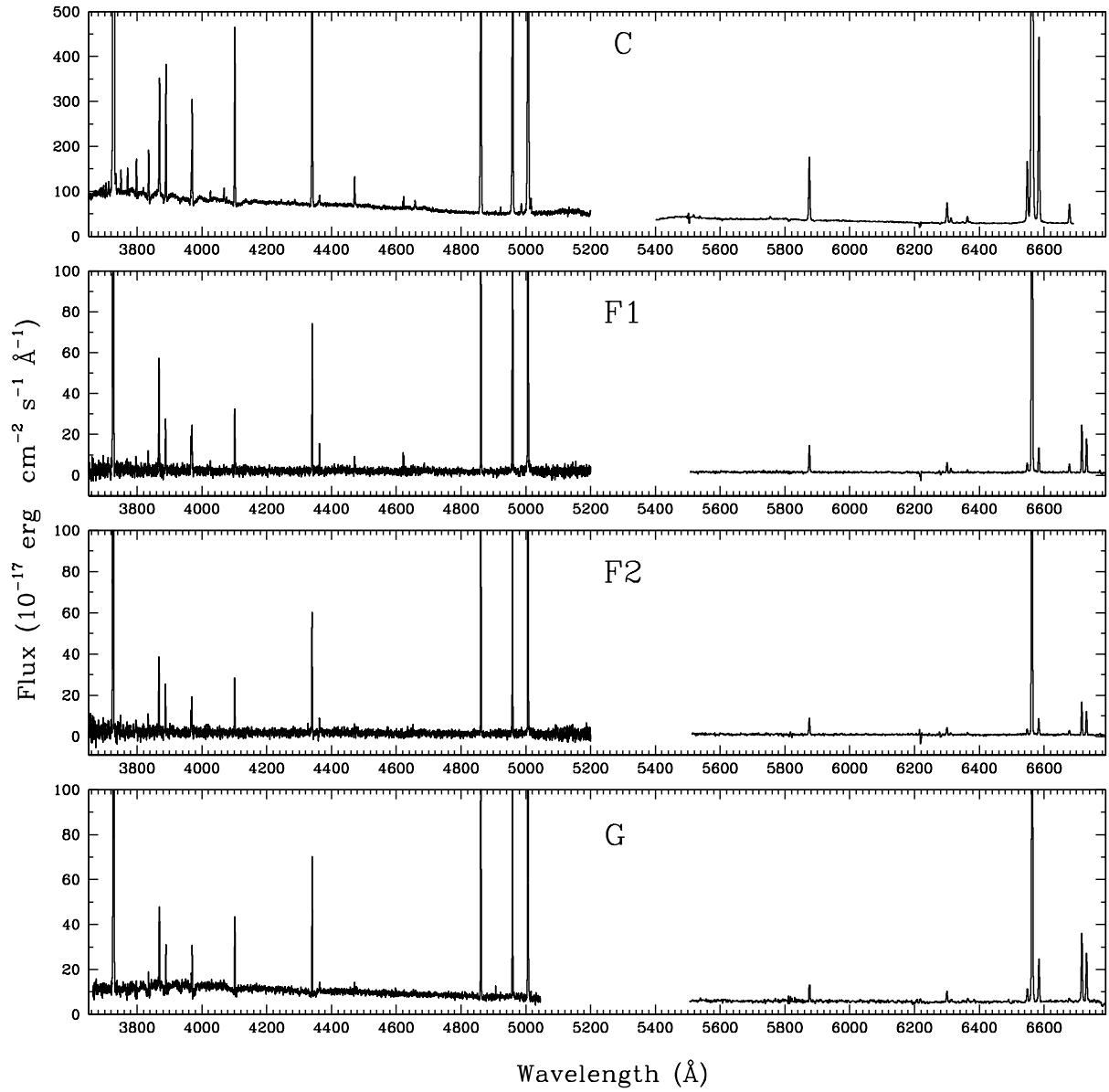}
\protect\caption[ ]{\small{ISIS WHT spectra of members C, F1, F2 and G. The spectra have been scaled down in flux in 
order to distinguish the faint lines.}}
\label{fig4}
\end{figure*}

In Figure~\ref{fig4} we show the wavelength and flux calibrated spectra of
members C, F1, F2, and G. They show the usual optical emission lines from
[\ion{O}{2}] $\lambda\lambda$3726,3729 to [\ion{S}{2}]
$\lambda\lambda$6717,6731. The  [\ion{O}{3}] $\lambda\lambda$4959, 5007  lines
are very bright relative to H$\beta$, reflecting the  high-excitation of the
ionized gas. We can also observe the rather high signal-to-noise ratio of the
[\ion{O}{3}]  $\lambda$4363 emission line in all these spectra, that will
allow us to obtain a good direct determination of the electron  temperature.
Member C is the most luminous galaxy of HCG~31, and it has the spectrum with
the highest signal-to-noise  ratio. Members B and G show some underlying
stellar absorption in their spectra, which is especially intense in B.  Galaxy H
is the faintest object detected. It could correspond to one of the SSC found by
\citet{JC00} in the tidal debris of the southwest tail.


The line intensities for each spectrum were measured by integrating all the
flux in the line between two given limits and  over a fitted local continuum.
The observed line intensities must be  corrected for
interstellar reddening. To do that, we have used the reddening constant,
C(H$\beta$), obtained from the  intensities of \ion{H}{1} Balmer lines in our
optical spectra for each burst. To obtain accurate values of the fluxes  of
nebular Balmer lines we have also to correct for the underlying stellar
absorption. Absorption wings were only  evident in H$\alpha$, H$\beta$ and
H$\gamma$ in the spectra of members B and G. The underlying stellar 
absorption is marginal for the rest of the objects. Following \citet{MB93}, we
have corrected for  absorption making use of the following relation: 
\begin{eqnarray}
I_{cor-abs}(\lambda) = I_0(\lambda) \frac{1+W_{abs}/W_{\lambda}}{1+W_{abs}/W_{H\beta}},
\end{eqnarray}
where $I_0(\lambda)$ and $I_{cor-abs}(\lambda)$ are the observed and corrected
fluxes and $W_{abs},\ W_{\lambda}$ and  $W_{H\beta}$ are the equivalent widths
of the underlying stellar absorption of the studied emission line, the 
equivalent width of the emission line, and the equivalent width of H$\beta$,
respectively. The value of $W_{abs}$ can change with the age of the burst, but
following \citet{O95}, we have adopted $W_{abs}=2$ \AA\ for all the observed
members in HCG~31. This is the same value adopted by \citet{R03}. Finally, the 
spectra were corrected for reddening using the \citet{W58} law and the value of
C(H$\beta$) derived from the comparison of the H$\beta$, H$\gamma$, and
H$\alpha$ relative intensities  with the theoretical values expected for case B
recombination using \citet{B71}. These theoretical ratios  were
I(H$\alpha$)/I(H$\beta$)=2.86 and I(H$\gamma$)/I(H$\beta$)=0.468, appropriate
for an electron temperature of  10$^4$ K and electron densities around 100
cm$^{-3}$. The reddening coefficient C(H$\beta$), the H$\beta$ line flux, 
$F(H\beta)$ (corrected by reddening and underlying stellar absorption) and the
equivalent width of several lines  [$W$(H$\beta$) and $W$([\ion{O}{3}])] are given
in Table~\ref{table3}, while the reddening-corrected line intensity ratios
relative to H$\beta$ of each member of HCG~31 are given in Table~\ref{table4} 
(H$\alpha$, H$\gamma$ and H$\delta$ intensity ratios are also corrected by underlying stellar absorption).
We have estimated  the error in the line intensities following the equation
given by \citet{Ca00}. Colons indicate errors of the order or  greater than
40\%.


\begin{table*}[t]
  \caption{General properties of bursts in HCG~31}
  \label{table3}
  \scriptsize  
  \begin{tabular}{lcccccccc}
    \\
    \tableline\tableline
	\noalign{\smallskip}
       &   A1  &  B &  C &  E  &  F1  &  F2 & G & H \\
    \tableline
 	\noalign{\smallskip}
   $-M_B$&15.75$\pm$0.06&18.71$\pm$0.05&19.43$\pm$0.05&15.76$\pm$0.06&15.76$\pm$0.06&14.34$\pm$0.06& 
18.88$\pm$0.05&$>$13.1\\
Slit (") &  3.6$\times$1 &   7.2$\times$1 & 12.96$\times$1 & 7.2$\times$1 & 7.2$\times$1 & 7.2$\times$1 & 3.6$\times$1 
& 3.6$\times$1 \\
    C(H$\beta$)&0.16$\pm$0.06& 
0.28$\pm$0.08&0.09$\pm$0.03&0.11$\pm$0.05&0.32$\pm$0.06&0.14$\pm$0.05&0.09$\pm$0.05&0.15$\pm$0.06\\
    $T_e[$O III$]$ (K)&8300\tablenotemark{a}&11500$\pm$700&9400$\pm$600&11100$\pm$1000&12600$\pm$1400&12300$\pm$1500  
&11600$\pm$700&9000\tablenotemark{a}\\    
$T_e[$O II$]$\tablenotemark{b} 
(K)&9900\tablenotemark{a}&12000$\pm$400&10800$\pm$300&11800$\pm$600&12600$\pm$700&12400$\pm$800&12000 
$\pm$400&10500\tablenotemark{a} \\
    $T_e[$N II$]$ (K)&\nodata&\nodata&10800$\pm$500&\nodata&\nodata&\nodata&\nodata&\nodata\\
    $N_e$ (cm$^{-3}$)& $<$100 & $<$100  &  210$\pm$70 & $<$100 & $<$100&$<$100 & $<$100&$<$100 \\
    $\Delta$v$_r$\tablenotemark{c}  &105$\pm$15 &132$\pm$25&   0 & $-$39$\pm$15 & $-$66$\pm$15 & $-$62$\pm$15& 
$-$2$\pm$15& 28$\pm$20\\
    F(H$\beta$)\tablenotemark{d}&2.90$\pm$0.23&17.8$\pm$0.5&461.3$\pm$1.8&15.7$\pm$0.6&31.87$\pm$0.16&             
22.76$\pm$1.86&25.94$\pm$0.59&1.44$\pm$0.16\\
    W(H$\beta$) (\AA) &27.0 $\pm$2.6&12.9$\pm$0.5&91.1$\pm$2.1& 
21.1$\pm$1.1&218$\pm$13&256$\pm$43&37.0$\pm$1.6&117$\pm$30\\
    W($[$O III$]$) (\AA)&46.2$\pm$4.4&26.7$\pm$1.2&218$\pm$4&61.0$\pm$1.5&1430$\pm$150&1192$\pm$135&104$\pm$10              
&347$\pm55$ \\
  \noalign{\smallskip}
  \tableline\tableline
  \end{tabular}
  \tablenotetext{a}{Calculated using empirical calibrations of \citet{P01a,P01b} and \citet{D02}.}
  \tablenotetext{b}{$T_e[$O II$]$ calculated from the relation given by \citet{S90}.}
  \tablenotetext{c}{Radial velocity with respect to member C, in km s$^{-1}$.}
  \tablenotetext{d}{In units of 10$\rm^{-16}$ erg s$\rm^{-1}$ cm$\rm^{-2}$.}
\end{table*}

\begin{table*}
  \caption{Dereddened line intensity ratios with respect to I(H$\beta$)=100.}
  \label{table4}  
  \scriptsize
  \begin{tabular}{lrcccccccc}
  \noalign{\smallskip}
    \tableline\tableline
	\noalign{\smallskip}
    Line & f($\lambda$)& A1 &B &  C &  E  &  F1  &  F2 & G & H \\
	\tableline    
	\noalign{\smallskip}
 3687 H 19	   &0.27&\nodata&\nodata&0.64:&\nodata&\nodata&\nodata&\nodata&\nodata\\
 3692 H 18	   &0.27&\nodata&\nodata&0.85:&\nodata&\nodata&\nodata&\nodata&\nodata\\
 3697 H 17	   &0.27&\nodata&\nodata&1.34$\pm$0.29&\nodata&\nodata&\nodata&\nodata&\nodata\\
 3704 HeI +H 16	   &0.26&\nodata&\nodata&1.49$\pm$0.30&\nodata&\nodata&\nodata&\nodata&\nodata\\
 3712 H 15	   &0.26&\nodata&\nodata&1.39$\pm$0.29&\nodata&\nodata&\nodata&\nodata&\nodata\\
 3726 $[$O II$]$    
&0.26&150$\pm$25&189$\pm$18&109.1$\pm$3.8&131$\pm$11&64.5$\pm$4.3&84$\pm$12&171$\pm$11&214$\pm$46\\
 3729 $[$O II$]$    
&0.26&214$\pm$33&281$\pm$24&123.1$\pm$4.1&217$\pm$17&96.7$\pm$5.6&126$\pm$16&240$\pm$15&295$\pm$56\\
 3734 H 13	   &0.26&\nodata&\nodata&2.13$\pm$0.25&\nodata&\nodata&\nodata&\nodata&\nodata\\
 3750 H 12	   &0.26&\nodata&\nodata&2.36$\pm$0.27&\nodata&3.4:&4.2:&\nodata&\nodata\\
 3771 H 11	   &0.26&\nodata&\nodata&2.92$\pm$0.28&\nodata&4.0:&4.9:&0.33:&\nodata\\
 3798 H 10	   &0.26&\nodata&5.8:&4.40$\pm$0.40&\nodata&4.6:&6.9:&0.88:&\nodata\\
 3819 He I	   &0.25&\nodata&\nodata&0.50:&\nodata&\nodata&\nodata&\nodata&\nodata\\
 3835 H 9	   &0.24&\nodata&7.1:&6.13$\pm$0.53&\nodata&7.2$\pm$3.5&8.2:&2.6:&\nodata\\
 3869 $[$Ne III$]$  &0.23&17.6:&31.8$\pm$6.5&15.09$\pm$0.87&35.1$\pm$9.3&39.4$\pm$9.7&32$\pm$11&28.6$\pm$3.7&26.1:\\
 3889 He I	   &0.22&13.7:&17.5$\pm$6.1&16.73$\pm$0.92&16.6$\pm$6.6&21.0$\pm$6.0&21.8$\pm$9.5&14.7$\pm$2.2& \nodata\\
 3968 $[$Ne III$]$  &0.21&\nodata&3.9$\pm$1.1&5.11$\pm$0.81&5.2$\pm$1.8&10.3$\pm$1.2&8.4$\pm$1.7&5.9$\pm$1.5&12.3:\\
 3970 H$\epsilon$  
&0.21&13.8:&13.2$\pm$1.9&4.49$\pm$0.74&7.8$\pm$1.5&14.5$\pm$1.3&13.3$\pm$2.2&12.4$\pm$1.9:&\nodata\\
 4026 He I	   &0.20&\nodata&\nodata&1.14$\pm$0.20&\nodata&1.8:&\nodata&\nodata&\nodata\\
 4068 $[$S II$]$    &0.19&\nodata&\nodata&1.85$\pm$0.21&6.4:&3.1$\pm$1.7&\nodata&3.2:&\nodata\\
 4076 $[$S II$]$    &0.19&\nodata&\nodata&0.54$\pm$0.16&\nodata&\nodata&\nodata&\nodata&\nodata\\
 4101 H$\delta$\tablenotemark{a}     
&0.18&23$\pm$11&23.8$\pm$2.5&23.2$\pm$0.8&25.3$\pm$6.0&25.0$\pm$4.5&24.7$\pm$8.3&21.2$\pm$2.1&19.4:\\
 4244 $[$Fe II$]$   &0.15&\nodata&\nodata&0.40:&\nodata&\nodata&\nodata&\nodata&\nodata\\
 4266 C II	   &0.15&\nodata&\nodata&0.22:&\nodata&\nodata&\nodata&\nodata&\nodata\\
 4287 $[$Fe II$]$   &0.14&\nodata&2.5:&0.78$\pm$0.12&\nodata&\nodata&\nodata&\nodata&\nodata\\
 4340 H$\gamma$\tablenotemark{a}    
&0.135&47$\pm$15&46.9$\pm$5.5&46.9$\pm$1.0&46.9$\pm$5.5&46.9$\pm$4.6&46.9$\pm$9.5&46.9$\pm$2.3&47$\pm$23\\
 4359 $[$Fe II$]$   &0.13&\nodata&\nodata&0.53$\pm$0.12&\nodata&\nodata&\nodata&\nodata&\nodata\\
 4363 $[$O III$]$   &0.13&2.8:&2.4$\pm$1.3&1.22$\pm$0.12&2.44$\pm$0.94&6.96$\pm$0.66&4.9$\pm$1.3&3.1$\pm$1.8&\nodata\\
 4471 He I	   &0.10&5.9:&3.6:&3.82$\pm$0.24&\nodata&4.8$\pm$1.7&4.5:&6.0:&\nodata\\
 4658 $[$Fe III$]$  &0.05&\nodata&2.5:&1.23$\pm$0.13&4.1:&\nodata&\nodata&2.0:&\nodata\\
 4713 He I	   &0.05&\nodata&\nodata&0.44$\pm$0.10&\nodata&\nodata&\nodata&\nodata&\nodata\\
 4861 H$\beta$     &0.00&100$\pm$16&100$\pm$5&100$\pm$1&100$\pm$5&100$\pm$2&100$\pm$16&100$\pm$5&100$\pm$35\\
 4881 $[$Fe III$]$  &0.00&\nodata&\nodata&0.35$\pm$0.10&2.3:&\nodata&\nodata&0.42:&\nodata\\
 4922 He I	  &-0.01&\nodata&\nodata&0.75$\pm$0.09&3.1:&\nodata&1.8:&1.4:&\nodata\\
 4959 $[$O III$]$  &-0.02&64$\pm$15&80.3$\pm$4.5&77.5$\pm$0.7&90$\pm$12&184$\pm$12&137$\pm$25&103$\pm$3&52.2:\\
 4986 $[$Fe III$]$ &-0.02&\nodata&7.3$\pm$1.7&1.39$\pm$0.15&4.9:&1.9:&\nodata&1.6:&\nodata\\
 5007 $[$O III$]$  &-0.03&197$\pm$32&243$\pm$10&232$\pm$2&274$\pm$38&546$\pm$23&413$\pm$47&306$\pm$7&136$\pm$28\\
 5016 He I	  &-0.03&\nodata&4.8:&2.22$\pm$0.19&\nodata&1.7:&2.7:&2.3:&\nodata\\
 5161 $[$Fe II$]$  &-0.04&\nodata&\nodata&0.43:&\nodata&\nodata&\nodata&\nodata&\nodata\\
 5200 $[$N I$]$    &-0.05&\nodata&\nodata&1.34$\pm$0.21&\nodata&\nodata&\nodata&\nodata&\nodata\\
 5518 $[$Cl III$]$ &-0.17&\nodata&\nodata&0.30:&\nodata&\nodata&\nodata&\nodata&\nodata\\
 5538 $[$Cl III$]$ &-0.18&\nodata&\nodata&0.33:&\nodata&\nodata&\nodata&1.8:&\nodata\\
 5755 $[$N II$]$   &-0.21&\nodata&\nodata&0.49$\pm$0.05&\nodata&\nodata&\nodata&\nodata&\nodata\\
 5876 HeI	  
&-0.23&14.2$\pm$2.5&13.6$\pm$1.5&11.2$\pm$0.4&9.46$\pm$0.82&11.0$\pm$2.6&11.0$\pm$1.1&12.0$\pm$0.9                      
&13.8:\\
 5980 Si II	  &-0.25&\nodata&\nodata&0.21:&\nodata&\nodata&\nodata&\nodata&\nodata\\
 6300 $[$O I$]$    
&-0.30&21.0$\pm$2.7&12.0$\pm$0.6&3.53$\pm$0.10&8.19$\pm$0.45&3.9$\pm$1.1&5.5$\pm$1.9&7.86$\pm$0.60                  
&15.4:\\
 6312 $[$S III$]$  &-0.30&\nodata&0.67:&0.86$\pm$0.03&1.3:&1.8:&1.5:&2.2:&\nodata\\
 6347 Si II	  &-0.31&\nodata&\nodata&0.18:&\nodata&\nodata&\nodata&\nodata&\nodata\\
 6364 $[$O I$]$    &-0.31&2.1:&4.9:&1.12$\pm$0.03&2.6:&1.3:&1.1:&\nodata&\nodata\\
 6371 Si II	  &-0.31&\nodata&\nodata&0.22:&\nodata&\nodata&\nodata&\nodata&\nodata\\
 6548 $[$N II$]$   
&-0.34&17.0$\pm$1.8&8.54$\pm$0.85&10.9$\pm$0.1&10.6$\pm$0.3&4.54$\pm$0.31&3.91$\pm$0.45&9.2$\pm$1.5&18.5:\\
 6563 H$\alpha$\tablenotemark{a}   
&-0.34&286$\pm$9&286$\pm$2&286$\pm$1&286$\pm$2&286$\pm$3&286$\pm$6&286$\pm$3&286$\pm$76\\
 6584 $[$N II$]$   
&-0.34&36.0$\pm$2.4&28.9$\pm$0.9&29.7$\pm$0.2&25.6$\pm$0.7&9.82$\pm$0.38&10.2$\pm$0.7&30.2$\pm$1.7&                      
30$\pm$14\\
 6678 He I	  &-0.35&\nodata&3.0:&3.12$\pm$0.03&2.38$\pm$0.37&3.33$\pm$0.17&3.39$\pm$0.38&\nodata&\nodata\\
 6716 $[$S II$]$   &-0.36&\nodata&14.1$\pm$0.3&\nodata&42.2$\pm$0.3&17.1$\pm$0.3&19.2$\pm$0.4&47.4$\pm$0.6&56$\pm$5\\
 6731 $[$S II$]$ &-0.36&\nodata&10.1$\pm$0.3&\nodata&29.6$\pm$0.4&11.8$\pm$0.2&14.0$\pm$0.4&33.4$\pm$0.6&40$\pm$4\\
 \noalign{\smallskip}
    \tableline\tableline
  \end{tabular}
   \tablenotetext{a}{Corrected by underlying stellar absorption assuming $W_{abs}$=2 \AA.}
\end{table*}


\subsubsection{Physical conditions of the ionized gas}

\begin{figure*}[h!]
\centering
\includegraphics[width=0.9\linewidth]{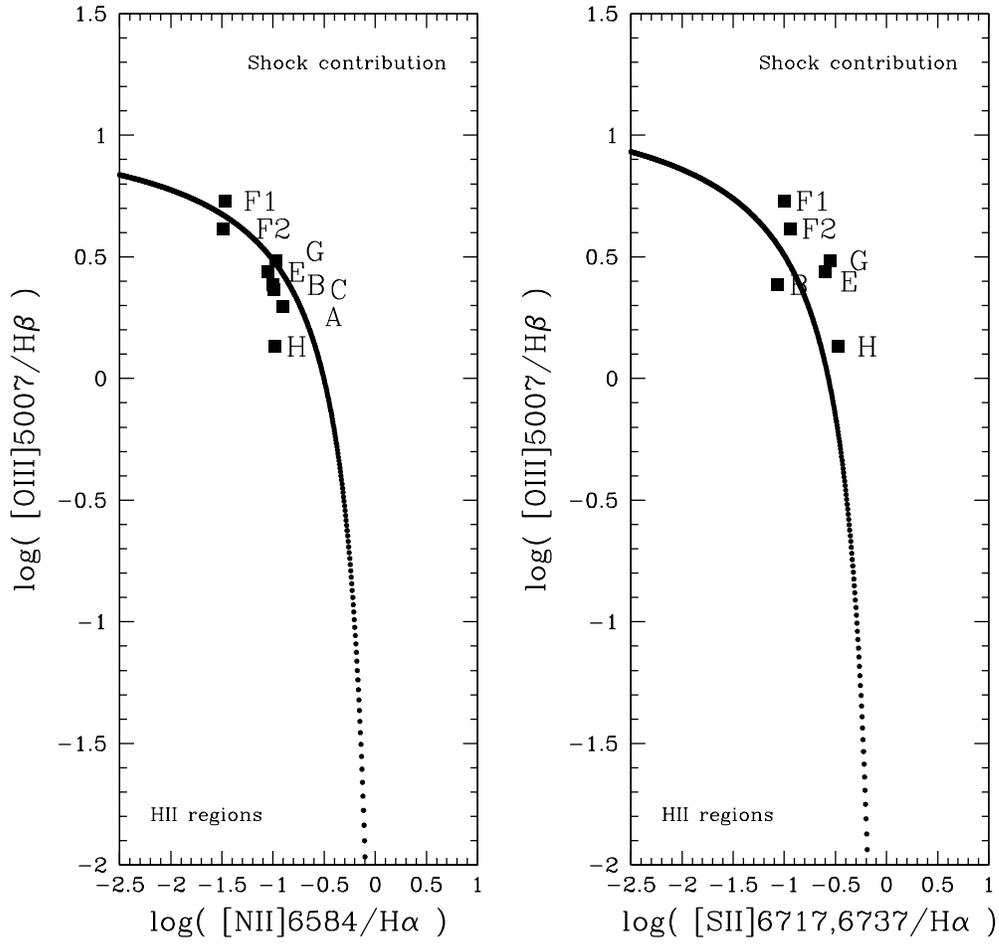}
\protect\caption[ ]{\small{Physical properties of the emission line regions in
HCG~31. The flux line ratios are indicated. The solid lines give the limit for
ionization by a zero age starburst, following \citet{Do00}.}}
\label{fig5}
\end{figure*}

\citet{VO87} and \citet{Do00} proposed diagnostic diagrams based on emission
line ratios useful for classifying the excitation mechanism of an \ion{H}{2}
region. These diagrams provide an objective  spectroscopical separation of
\ion{H}{2} regions from other classes of narrow line regions associated with 
active galactic nuclei. We have plotted these two relations in
Figure~\ref{fig5} together with our observational data.  From the figure, it is
evident that our data are consistent with the \emph{loci} of typical \ion{H}{2}
regions.


As it can be seen in Table~\ref{table4}, we observe the auroral [\ion{O}{3}]
$\lambda$4363 emission line in most of  the objects: B, C, E, F1, F2, and G.
Therefore, we can determine the electron temperature, $T_e$, using the 
[\ion{O}{3}] ($\lambda$4959+$\lambda$5007)/$\lambda$4363 ratio for those
objects. In the case of member C, we also  measure the [\ion{N}{2}]
$\lambda$5755 line and we can obtain an  additional direct determination of
$T_e$ for this ion. Electron temperatures are calculated making use of the 
five-level program for the analysis of emission-line nebulae that is included
in IRAF NEBULAR task \citep{SD95}. It is  usual to use a two-zone approximation
to define the temperature structure of the nebulae, assuming
$T_e$([\ion{O}{3}]) as  the representative temperature for high ionization
potential ions and $T_e$([\ion{N}{2}]) or $T_e$([\ion{O}{2}]) for the low
ionization potential ones. In our case, we have only determined both
temperatures in the case of member C. For the rest of  objects, we have assumed
the relation between $T_e$([\ion{O}{3}]) and $T_e$([\ion{O}{2}]) based on
the photoionization models by  \citet{S90}:
\begin{eqnarray}
t(\mbox{[\ion{O}{2}]}) = \frac{2}{t(\mbox{[\ion{O}{3}]})^{-1} + 0.8},
\end{eqnarray}
where $t=T_e/10^4$. It is remarkable that in the case of member C, the direct
value of $T_e$([\ion{N}{2}]) and the  $T_e$([\ion{O}{2}])  estimated using (2)
coincide (see Table~\ref{table3}). Note that with the alternative linear relation
between $T_e$([\ion{O}{3}]) and $T_e$([\ion{O}{2}]) given by \citet{G92}, the
values of $T_e$([\ion{O}{2}]) would be around 1000 K lower.  In the cases of
members A1 and H we do not have direct  determinations of the electron
temperature. In these cases, the adopted values of $T_e$ are those that
reproduce the  chemical abundances obtained for these zones applying empirical
methods (see \S~3.3.2).

All the members of HCG~31 show the [\ion{O}{2}] $\lambda\lambda$3726, 3729
doublet and the [\ion{S}{2}] $\lambda\lambda$6716, 6731 doublet (except
members A1 and C, for which \ion{S}{2} lines are out of the observed range). 
We have used both
doublets to derive the electron  density, $N_e$, of the ionized gas.  Electron
densities are always below the low-density limit ($<$100 cm$^{-3}$) except for
member C, where $N_e$ is  slightly higher (210 cm$^{-3}$). The final adopted
values of $T_e$ and $N_e$ (or its upper limit) are compiled in 
Table~\ref{table3}.


\subsubsection{Abundance Analysis}

Once the electron density and temperature are adopted for each burst, the ionic
abundances of He$^+$, O$^+$,  O$^{++}$, N$^+$, Ne$^{++}$, S$^+$, S$^{++}$ and
Fe$^{++}$ can be derived from our spectra. In the case of  O$^+$, O$^{++}$,
N$^+$, Ne$^{++}$, S$^+$, and S$^{++}$ we have used the IRAF NEBULAR task to
derive their ionic  abundances from the intensity of collisionally excited
lines.  We have assumed a two-zone scheme for deriving the ionic abundances,
adopting $T_e$([\ion{O}{3}]) for the high ionization  potential ions O$^{++}$,
Ne$^{++}$, and S$^{++}$; and $T_e$([\ion{N}{2}]) or $T_e$([\ion{O}{2}]) for the
low ionization  potential ions O$^+$,  N$^+$, S$^+$ and Fe$^{++}$. These ionic
abundances are listed in Table~\ref{table5}. The electron density assumed is 
100 cm$^{-3}$ for all the members except C, for which we have used 210
cm$^{-3}$. 

\begin{figure*}[h!]
\centering
\includegraphics[width=1\linewidth]{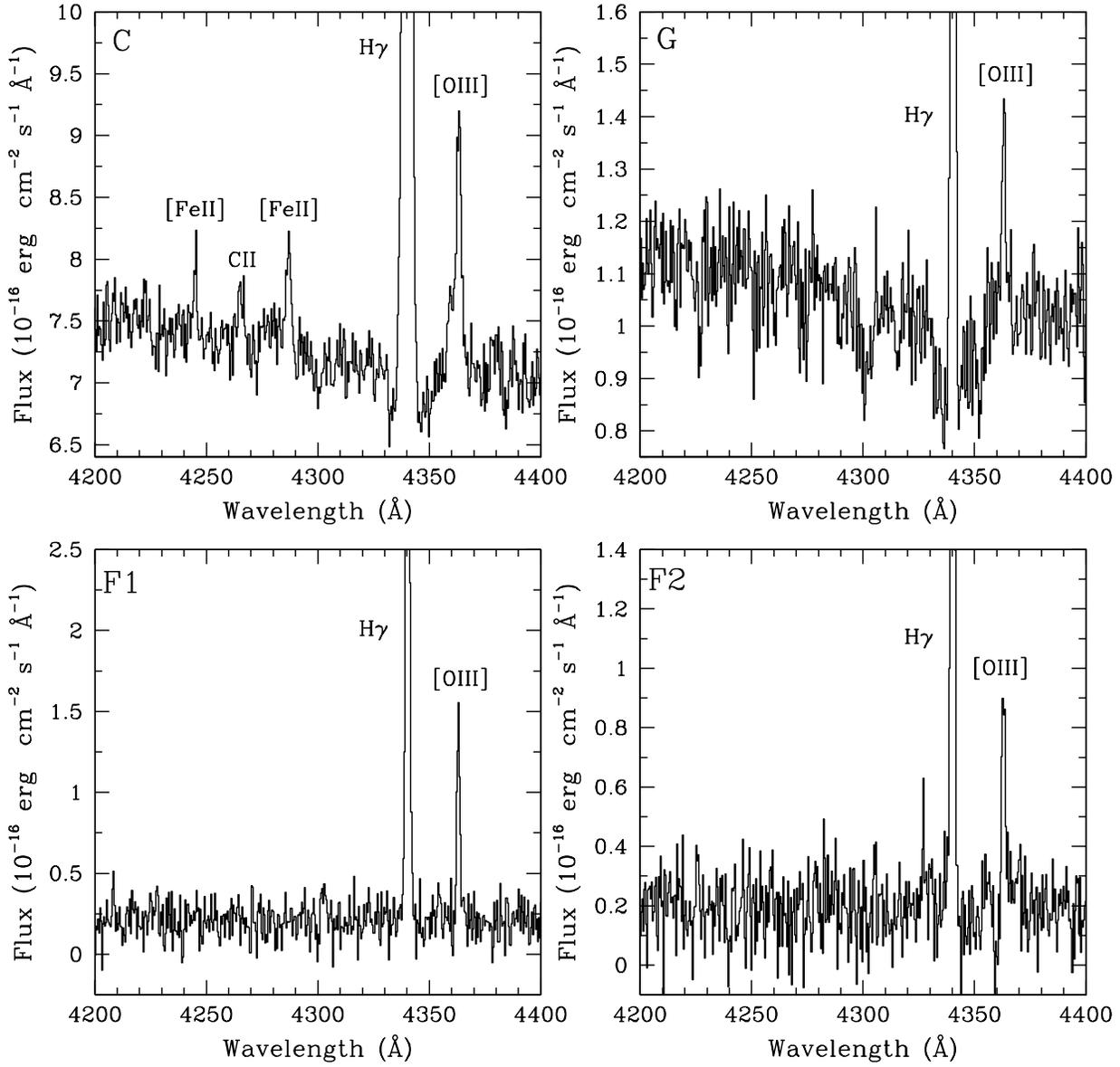}
\protect\caption[ ]{\small{Detail of the spectra of members C, F1, F2, and G
showing the zones around the [\ion{O}{3}]  $\lambda$4363 emission line. Member C
also shows the \ion{C}{2} $\lambda$4267 and [\ion{Fe}{2}] $\lambda\lambda$4244, 
4287 emission lines. H$\gamma$ shows a weak absorption due to the
underlying stellar population in members C  and G.}}
\label{fig6}
\end{figure*}

An interesting result is the possible detection of \ion{C}{2} $\lambda$4267 in
the spectrum of member C (see  Figure~\ref{fig6}). This is the first time this
recombination  line is reported in an \ion{H}{2} galaxy and opens new
possibilities for extragalactic C abundance studies to be done  in the future
with large aperture telescopes. Although the signal-to-noise ratio of
\ion{C}{2} $\lambda$4267  is  rather poor, we have calculated the corresponding
C$^{++}$/H$^+$ ratio. We have used the effective recombination  coefficients
calculated by \citet*{D00} and $T_e$([\ion{O}{3}]) as representative of the zone
where this ion is located. 

The He$^+$/H$^+$ ratio has been derived from the brightest \ion{He}{1} lines
observed in each object using the line emissivities calculated by
\citet*{SSM96}. We have corrected all the He$^+$/H$^+$ ratios for the
collisional  contribution following the calculations by \citet{KF95} and
\citet*{B02}. In Table~\ref{table5} we  present the He$^+$/H$^+$ ratios for
each individual \ion{He}{1} line and the final adopted average value for each 
object  $<$He$^+$/H$^+$$>$. 

Although we detect the [\ion{Fe}{2}] $\lambda$4244, $\lambda$4287, and
$\lambda$5161 emission lines in member C, the  Fe$^+$ abundance cannot be
calculated because these lines are affected by fluorescence \citep{Ve00}. The
Fe$^{++}$  abundances have been derived using a 34 level model atom that uses
the collision strengths calculated by \citet{Z96} and the transition 
probabilities given by \citet{Q96}. We have used the intensity ratio of the
[\ion{Fe}{3}] $\lambda\lambda$4658, 4986  lines to derive a further estimate
of N$_e$ in member C (the only region where the intensities measured for these 
lines have low uncertainties). We find N$_e$= 140$\pm$50 cm$^{-3}$, in
agreement with the value obtained from the  [\ion{O}{2}] diagnostic lines.

The total abundances have been determined for O, N, S, and Fe. The absence or
weakness of the \ion{He}{2} $\lambda$4686  line implies a negligible amount of
O$^{3+}$ in the nebula. Therefore, we can adopt the usual relation O/H = 
O$^+$/H$^+$ +  O$^{++}$/H$^+$. For N, we have assumed the standard ionization
correction factor by \citet{PC69}: N/O = N$^+$/O$^+$, which  is a reasonably
good approximation for an object with the excitation degree of the objects of
HCG~31. We have  measured two ionization stages of S in five of the objects.
Taking into account the relatively large ionization  degree of the spectra,
some contribution of S$^{3+}$ is expected. This ion does not show emission
lines in the optical  region. Therefore, a lower limit to the S abundance is
given. For Fe, the total abundances have been obtained from the  relation
\citep{RR03}: 
\begin{eqnarray}
\bigg[\frac{\rm{Fe}}{\rm{H}}\bigg] = \bigg[\frac{\rm{O}^+}{\rm{O}^{++}}\bigg]^{0.09} 
\bigg[\frac{\rm{Fe}^{++}}{\rm{O}^+}\bigg]\bigg[\frac{\rm{O}}{\rm{H}}\bigg].
\end{eqnarray}

The two objects with the best Fe/H determinations in HCG~31 show
$\log(\mbox{Fe}/\mbox{O})=-1.87$ (member B) and
$\log(\mbox{Fe}/\mbox{O})=-2.12$ (member C).  In a recent paper, \citet{Ro03}
has determined the Fe/O ratios in two giant \ion{H}{2} regions of the Local
Group with O abundances similar to those of our galaxies: 30 Dor in the LMC
($\log(\mbox{Fe}/\mbox{O})\simeq-2.3$) and N88A in the SMC
($\log(\mbox{Fe}/\mbox{O})\simeq-1.6$).  The Fe/O ratios of these giant
\ion{H}{2} regions (located in dwarf irregular star-forming galaxies) are
similar to the values we obtain for the galaxies of HCG~31. These values are
much lower than the solar abundance: $\log(\mbox{Fe}/\mbox{O})=-1.29$
\citep{Ho01}. The true Fe/O ratio in a given object depends on two factors: the
intrinsic value of Fe/O (in gas and dust), which depends on the star formation
history of the system, and the amount of Fe depleted in dust grains. Assuming a
solar value for the intrinsic Fe/O in our galaxies, we estimate that $\sim80\%$
of the Fe atoms are depleted onto dust grains.

For those objects without direct determination of $T_e$ (objects A1 and H) we
have derived the total O abundance  making use of empirical calibrations.
We have used the R$_{23}$ parameter:
\begin{eqnarray}
\rm{R_{23}}\equiv\frac{I([\textsc{O~ii}]\ 3727)+I([\textsc{O~iii}]\ 4959 + 5007)}{I(H\beta)},
\end{eqnarray}
and the excitation parameter, $P$:
\begin{eqnarray}
P\equiv\frac{I([\textsc{O~iii}]~4959 + 5007)}{I(H\beta)}\frac{1}{\rm{R_{23}}},
\end{eqnarray}
\citep{P01a,P01b} and the [\ion{N}{2}] $\lambda$6584/H$\alpha$
ratio \citep*{D02}. In  Table~\ref{table5} we also include the
values obtained with the empirical calibrations for all the members.
The objects in HCG~31 have O abundances that do not permit to discern between
the high metallicity  branch (calibrated by \citealt{P01a}) or the low
metallicity branch (calibrated by \citealt{P01b}). We have adopted the  average
value between both calibrations; this seems not to be a bad choice because
these values are rather similar  (to within 0.1 dex) to the ones obtained from
direct determination of $T_e$. In the other hand, the abundances obtained  from
the empirical calibrations of \citet{D02} are about 0.2 to 0.3 dex higher than
those derived from direct methods,  except for members F1 and F2 (the lowest
metallicity objects). Finally, we have assumed that the O/H abundance ratios 
of A1 and H are the average value between the three mentioned empirical
calibrations. Nevertheless, the true O/H  abundance ratios of A1 and H could be
somewhat lower than the values indicated in Table~\ref{table5}.


\begin{table*}[t]
  \caption{Chemical abundances of bursts in HCG~31}
  \label{table5}
  \scriptsize
  \begin{tabular}{lcccccccc}
  \noalign{\smallskip}
    \tableline\tableline
	\noalign{\smallskip}
                           &   A1  &  B   &  C   &  E   &  F1  &  F2  & G    & H   \\
    \tableline
    \noalign{\smallskip}
 12+log O$\rm^+$/H$\rm^+$&8.15\tablenotemark{a}&7.93$\pm$0.09&7.82$\pm$0.07&7.83$\pm$0.13&7.39$\pm$0.11&7.53$\pm$0.17& 
7.76$\pm$0.09&8.20\tablenotemark{a} \\
  
12+log O$\rm^{++}$/H$\rm^+$&8.15\tablenotemark{a} 
&7.73$\pm$0.08&8.00$\pm$0.04&7.83$\pm$0.07&7.96$\pm$0.06&7.87$\pm$0.08
&7.82$\pm$0.06&7.87\tablenotemark{a}\\ 
  
12+log O/H&8.35\tablenotemark{b}&8.14$\pm$0.09&8.22$\pm$0.05&8.13$\pm$0.10&8.07$\pm$0.07&8.03$\pm$0.11&8.15$\pm$0.08& 
8.37\tablenotemark{b}\\
  12+log O/H\tablenotemark{c}     &  8.22  & 8.22 & 8.15 & 8.18 & 8.12 & 8.13 & 8.18 & 8.33 \\
  12+log O/H\tablenotemark{d}    &  8.47  & 8.39 & 8.40 & 8.36 & 8.06 & 8.02 & 8.41 & 8.41 \\

$-$log(N/O)&1.21:&1.39$\pm$0.14&1.12$\pm$0.10&1.26$\pm$0.19&1.27$\pm$0.19&1.43$\pm$0.28& 1.31$\pm$0.13&1.33:\\
  12+log N/H        
&7.24:&6.74$\pm$0.23&7.10$\pm$0.15&6.87$\pm$0.29&6.79$\pm$0.26&6.61$\pm$0.37&6.83$\pm$0.20&7.04:\\
  log C$\rm^{++}$/O$\rm^{++}$  & \nodata & \nodata &  +0.32: & \nodata & \nodata & \nodata & \nodata & \nodata \\

12+log Ne$\rm^{++}$/H$\rm^+$&7.06:&7.31$\pm$0.19&7.29$\pm$0.17&7.41$\pm$0.27&7.16$\pm$0.25&                  
7.11$\pm$0.31&7.26$\pm$0.15&7.66:\\

  12+log S$\rm^+$/H$\rm^+$&\nodata&5.57$\pm$0.06&5.67$\pm$0.08\tablenotemark{e}&6.06$\pm$0.10&5.60$\pm$0.06&         
5.67$\pm$0.09&6.09$\pm$0.05&6.30\\
  12+log S$\rm^{++}$/H$\rm^+$&\nodata&5.97:&6.44$\pm$0.09&6.31:&6.25:&6.21: & 6.17: & \nodata \\
  12+log S/H   &\nodata&$>$6.16&$>$6.55&$>$6.55&$>$6.38&$>$6.36&$>$6.48& \nodata \\
  12+log He$\rm^+$/H$\rm^+$ (4471) & 11.08: & 10.87: & 10.88$\pm$0.04  &  \nodata & 10.99$\pm$0.22  & 10.96: & 11.09: 
& \nodata\\
  12+log He$\rm^+$/H$\rm^+$ (5876) & 11.01: & 11.01$\pm$0.10  & 10.90$\pm$0.03  & 10.85$\pm$0.11    & 10.92$\pm$0.20  
& 10.92$\pm$0.16  & 10.95$\pm$0.08  & 11.01:  \\
  12+log He$\rm^+$/H$\rm^+$ (6678) &\nodata & 10.90: & 10.91$\pm$0.03  & 10.80$\pm$0.14  & 10.95$\pm$0.12  & 
10.96$\pm$0.15  & \nodata& \nodata\\
  $<12+\rm{log}$ He$\rm^+$/H$\rm^+$$>$ & 11.01: & 11.01$\pm$0.10  & 10.90$\pm$0.03  & 10.83$\pm$0.13    & 
10.96$\pm$0.18  & 10.94$\pm$0.16  & 10.95$\pm$0.08  & 11.01: \\
  12+log Fe$\rm^{++}$/H$\rm^+$&\nodata&6.04$\pm$0.14 &5.71$\pm$0.09 & 6.06:  & 5.54:  & \nodata & 5.75: & \nodata\\
  12+log Fe/H                 &\nodata&6.27$\pm$0.23 &6.10$\pm$0.16 & 6.36:  & 6.16:  & \nodata & 6.14: & \nodata\\
  \tableline
  \noalign{\smallskip}
  Z/Z$_{\odot}$\tablenotemark{f}&0.46&0.28$\pm$0.07&0.34$\pm$0.04&0.28$\pm$0.07&0.24$\pm$0.04&              
0.22$\pm$0.06&0.29$\pm$0.06&0.48\\
  \tableline\tableline
  \end{tabular}
  \tablenotetext{a}{Calculated using $T_e$ estimated from empirical calibrations.}
  \tablenotetext{b}{Average of values obtained using the empirical calibrations of \citet{P01a,P01b} and \citet{D02}.}
  \tablenotetext{c}{Average of values obtained using the R$_{23}$ and $P$ parameters and the empirical calibration of 
\citet{P01a,P01b}.}
  \tablenotetext{d}{Determined using the [N II]/H$\alpha$ ratio \citep{D02}.}
  \tablenotetext{e}{Determined from the [S II] $\lambda\lambda$4068, 4076  emission lines.}
  \tablenotetext{f}{Asuming Z$_\odot=8.69\pm0.05$ \citep*{AP01}.}

\end{table*}


The O/H ratios obtained from direct determination of the electron temperatures
(all the objects except A1 and H), are  rather similar and range between 8.0
and 8.2 (in units of 12+log (O/H)). There are several previous determinations
of  the chemical abundances for some of the members of the group. For member A,
we have values of 8.30 \citep{R90}, 8.04  $\pm$ 0.06 \citep{IT98}, and 8.1
$\pm$ 0.2 \citep{R03}. These determinations are based  on temperatures obtained
from the measurement of the [\ion{O}{3}] $\lambda$4363 line, and are lower than
the value of O/H =  8.35  we obtain for A1 from empirical calibrations.
Therefore, and taking into account the systematically higher abundances given
by the empirical calibrations, we will adopt the value by \citet{IT98} as 
representative of member A. We prefer the determination of these authors
because of the high quality of their  spectrum. In the case of the brightest
object, member C, our value of 8.22 $\pm$ 0.05 is consistent with previous 
determinations by \citet{R90}, \citet{VC92} and \citet{R03}. \citet{R03} obtain
a direct determination of $T_e$  for  F1 (with a rather poor measurement of the
[\ion{O}{3}]\ $\lambda$4363 line) and determine O/H = 8.1 $\pm$ 0.2, which is 
very similar to our value of 8.07 $\pm$ 0.07. Finally, we obtain the first
direct O abundance determinations for  members B, E, F2, and G. The O/H ratios
implied by \citet{G92} linear relation between $T_e$([\ion{O}{3}]) and 
$T_e$([\ion{O}{2}]) would be slightly higher, by 0.01 dex (members F1 and F2)
and 0.10 dex (member C). The variation is almost negligible in the case of
N/H. These variations do not affect qualitatively to our conclusions.    

In the case of C, we have only an uncertain determination of C$^{++}$/H$^+$ for
member C. Following \citet{EP02} and  the photoionization models of \citet{S90}
and \citet{G99}, we can assume C$^{++}$/O$^{++}$ = C/O as representative for 
this object. We obtain log(C$^{++}$/O$^{++}$) = +0.32:, which is an extremely
high value. Values so high have never been reported for any extragalactic
object (see the compilation of \citealt{G03}).  There are several
determinations of the C$^{++}$/O$^{++}$ ratio obtained from the \ion{C}{2}
$\lambda$ 4267 line for  giant \ion{H}{2} regions of the Local Group.
\citet{EP02} have obtained values of log(C$^{++}$/O$^{++}$)=$-$0.19 and 
$-$0.39 for NGC~604 and  NGC~5461, respectively, and \citet{P03} derived
$-$0.50 for 30 Dor. These values correspond to the ratio of ionic  abundances
determined from \ion{C}{2} and \ion{O}{2} recombination lines, which give
systematically larger  abundances relative to H than the collisionally excited
lines. For these three reference objects: NGC~604, NGC~5461, and 30 Dor, we can
calculate the C$^{++}$/O$^{++}$ ratio obtained from the \ion{C}{2}
$\lambda$4267 and the  [\ion{O}{3}] lines (the valid quantity to compare with
the ratio we obtain for member C) and we get larger ratios 
(log(C$^{++}$/O$^{++}$)=+0.01 dex at most), but never as high as the ratio we
obtain for member C. A possible explanation is that we have simply overestimated the
intensity of the weak \ion{C}{2} $\lambda$4267 line by a factor of 2, but there
are three other possible explanations:

\begin{enumerate}
\item the \ion{C}{2} $\lambda$4267 line has been misidentified. This is
quite unlikely, because its wavelength  coincides exactly with the expected one
for that line. There is not any other observable nebular line (or stellar emission 
feature) at or very near ($\pm$1\AA) that wavelength, its width is similar to 
that of the other nebular lines of that spectral zone;
\item the line is abnormally intense due to an unknown exotic radiative
process;
\item the difference between abundances derived from collisionally excited and
recombination lines of the same ion  is especially large for this object (see
\citealt{E02} for a review on this problem in \ion{H}{2} regions).
\end{enumerate} 

If the last reason is the real one, it could be related to the presence of very
strong temperature fluctuations in  the ionized gas of member C. This
fluctuations would be stronger than in any other known \ion{H}{2} region to
account  for the extremely large value of the C$^{++}$/O$^{++}$ ratio and would
imply that the true metallicity of the object  should be much larger, and
closer to the solar value. 

\subsubsection{The Wolf-Rayet bump}

\begin{figure*}[h!]
\centering
\includegraphics[width=1\linewidth]{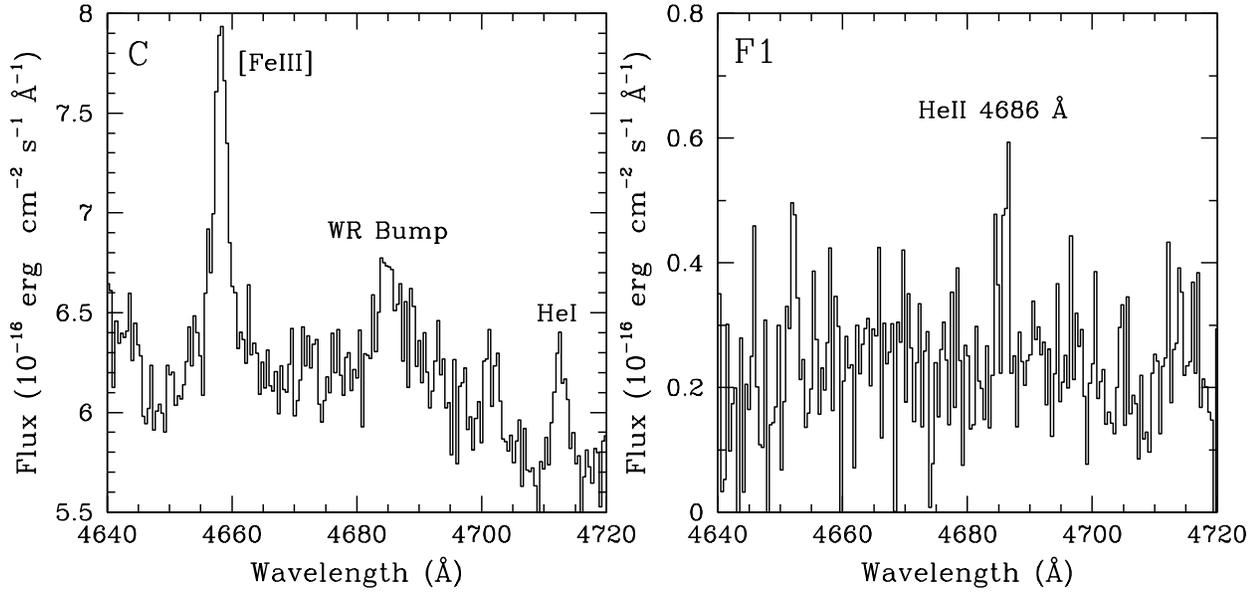}
\protect\caption[ ]{\small{Detail of the spectra of members C and F1 showing
the WR bump or the \ion{He}{2}  $\lambda$4686 emission. [\ion{Fe}{3}]
$\lambda$4658  and \ion{He}{1} $\lambda$4713 emission lines are also clearly 
observed in C.}}
\label{fig7}
\end{figure*}

The blue WR bump between 4650 and 4698 \AA\ can be clearly observed in member C
(see Fig.~\ref{fig7}). It is a  blend of emission features of He
(\ion{He}{2} $\lambda$4686), C (\ion{C}{3}/\ion{C}{4} $\lambda$4650), and N 
(\ion{N}{3} $\lambda$4640) of WR stars. The strength of these lines can be used
to estimate the number of WR stars in  a galaxy (e.g. \citealt{VC92},
\citealt{SV98}). In the spectrum of member C, the strongest contribution is
\ion{He}{2}  $\lambda$4686, although there is a nearby relatively bright
[\ion{Fe}{3}] line. \citet{KS86} reported the first  \ion{He}{2} $\lambda$4686
line in the spectrum of the main body of HCG~31, but they also noted the
\ion{N}{3}  $\lambda$4640 feature. \citet*{GIT00} detected  \ion{N}{3}
4512$\lambda$ and \ion{Si}{3} $\lambda$4565 emission lines, as well as the
\ion{N}{2} $\lambda$5720-40 one.  Despite a possible contribution of nebular
emission, these features are related to WN stars, although \ion{N}{2} 
$\lambda$5720-40 could be contaminated by other WR subtypes.  The observed
differences between our spectrum of member C and the previously reported ones
could be due because we  are not observing the same region inside the A+C
complex. \citet{GIT00} remarked than the WR  population in the main body of HCG
31 is dominated by WNL stars, because the \ion{C}{4} $\lambda$5808 \AA\
emission  feature (the red WR bump) was not detected in their spectrum. We also
do not detect that spectral feature in any of  our spectra.


We have detected a faint nebular \ion{He}{2} $\lambda$4686 emission line in
member F1 that has not been previously  reported. This feature seems also to be
present in F2. Moreover, a very faint blue WR bump seems to be present in the
spectra  of A1 and B.

\subsubsection{Kinematics of HCG~31}

We have determined the mean radial velocity of each member of HCG~31 from the
centroid of H$\beta$ and [\ion{O}{3}]  $\lambda$5007 emission lines in the
brightest zone of each galaxy. We have taken the most luminous member, C, as
reference for the  radial velocities of the group; its heliocentric velocity is
4042 km s$^{-1}$. The relative radial velocities of the  group members are
given in Table~\ref{table3}. They span a narrow interval of values with a
maximum difference of  about 200 km s$^{-1}$. Our velocities are in very good
agreement with those obtained by \citet{R90} and \citet{R03}  from their
optical spectra and those obtained by \citet{W91} from their \ion{H}{1}
velocity map. We find  that members E, F1, F2, and G, all of them located to
the southern part of  the system, show negative values of the velocity. This is
consistent with the general trend observed in the \ion{H}{1} velocity map of
\citet{W91}. However,  the new object H has a somewhat abnormal positive
velocity relative to its position. The other northern  members of the group, A1
and B, show positive velocities. 

\begin{figure*}[ht!]
\centering
\includegraphics[width=1\linewidth]{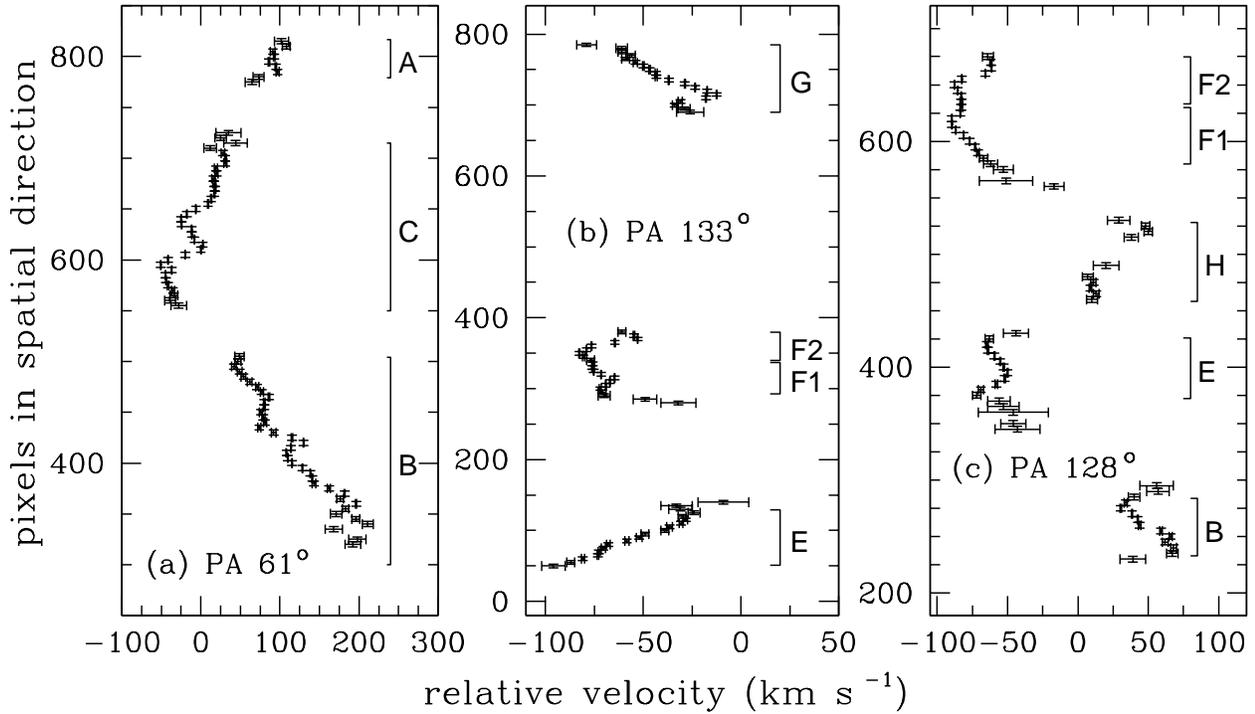}
\protect\caption[ ]{\small{a, b and c: Position-velocity diagrams for the three
slit position observed in HCG~31 analyzed in 1\arcsec\ bins. The horizontal
bars represent the uncertainty of the Gaussian fitting for each point. The 
location and extension of the different galaxy members is also indicated. North
is up in (a) and southeast is up in (b) and (c).}}
\label{fig8}
\end{figure*}

We have also studied the kinematics of the ionized gas via the spatially
resolved analysis of bright emission line  profiles along each  slit position.
We have extracted zones of 5 pixels (1\arcsec\ in the spatial direction)
covering all the extension of the line  emission in the three slit positions
shown in Figure~\ref{fig3}. This analysis has been performed via Gaussian fitting
making  use of the Starlink DIPSO software. For each zone, 
we have analyzed the profiles of H$\beta$
and the [\ion{O}{3}] $\lambda\lambda$4959, 5007 lines (the brightest lines of
the spectra taken with the blue arm of the spectrograph, with the highest
spectral resolution) using one, two or all of these lines depending on the
signal-to-noise of  each spectrum. The final adopted values are the average of
the results for all the individual lines used in the  analysis of each
individual zone. In all the cases, the line emission profiles were well fitted
by a single Gaussian  fit and therefore  no complex profiles were needed. In
Figure~\ref{fig8} we show the position-velocity diagrams for the three slit 
positions. All  the velocities are referred to the mean heliocentric velocity
of member C (4042 km s$^{-1}$). The position of the  different galaxy members
of the group is also indicated on Figure~\ref{fig8}. \citet{R90} also show
position-velocity  diagrams for A, B, C, and E (extracted every 3.5\arcsec\
along the slit) and are very similar to ours in the zones in  common.
\citet{R03} obtain Fabry-Perot observations of the group in H$\alpha$ and also
present  position-velocity diagrams for A, B, and C at the same positions,
angles, and spatial resolution as \citet{R90}. 


Our position-velocity diagram for C, as well as that obtained by \citet{R90},
shows a sinusoidal pattern in  the center of the object. It is evident that the
velocity reverses in the central region with countermotions of the  order of 50
km s$^{-1}$. This is a traditional diagnostic of interaction and merging of
galaxies (e.g. \citealt{S82,  R90}) and indicates that a merging process is
ongoing in the central zone of the A+C complex. However, \citet{R03} find  that
members A and C seem to be a single kinematical entity in their fabry-Perot
data.  Perhaps, \citeauthor{R03} do not observe this feature due to the lower
sensitivity and lower  spatial resolution of  their position-velocity
diagrams. 

In the case of member B, our results are consistent with those from other
authors. Its variation of velocity across the slit is  rather linear indicating 
that the galaxy is basically  rotating as a solid-body, although some slight 
deviations from that (just concentrated in the nucleus)
are present.  There is a very faint nebular emission between galaxies A and B
for which it was not possible to carry out the  profile analysis. However, the
adjacent external zones of both galaxies seem to indicate a continuous linear 
connection in velocity of the ionized gas between both objects. The solid-body
rotation pattern of member B could be  affected by a possible tidal streaming
motion between B and the A+C complex, at least in the external zones. As in 
the case of member B, the ionized gas between zone A1 and the northeast
extension of member C seem to follow a smooth  linear increase of radial
velocity.

The position-velocity diagrams for PA 133$^\circ$ and PA 128$^\circ$ are rather
complex. The slit position at PA  133$^\circ$ (Fig.~\ref{fig8}b) was intended
to cover precisely the major axis of F1 and F2, and crosses the northern 
diffuse outskirts of member E and the center of member G. The slit position at
PA 128$^\circ$ (Fig.~\ref{fig8}c)  covers the main body of E, the zone around
H, and the north of F1 and F2. In Figure~\ref{fig8}b it can be seen that  the
diffuse faint area at the north of member E (indicated as ``E'') shows a linear
variation with a quite wide  amplitude of about 70 km s$^{-1}$. This behavior is
quite different to that shown by the main body of object E 
(Fig.~\ref{fig8}c), which, interestingly, does not show traces of the linear
behavior of the diffuse gas at the  north. In fact, the central part of E shows
a narrow sinusoidal behavior with a velocity amplitude of about 30--40 
km s$^{-1}$. This suggests that perhaps we are seeing two different
kinematical objects that coexist in  apparent close proximity. As it can be
seen in the $R$-band image shown in Figure~\ref{fig1} (as well as in the
other  optical images) the diffuse gas to the north of E delineates a tail
emerging from the A+C complex. Although member E  seems to be part of that tail
due to its location, its different kinematics suggests that perhaps it does not
form  part (at least kinematically) of it. On the other hand, from the images
and the results in Figure~\ref{fig8}, it is  apparent that member H seems
kinematically to be an extension of the aforementioned tail. If we inspect the
velocity  pattern of H in Figure~\ref{fig8}c, we can see that its peculiar
positive velocity can be interpreted as an  extension to slightly higher
positive velocities of the velocity distribution we see in PA 133$^\circ$ at
the north of  E. These indications suggest that the faint tidal tail emerging
from the southwest of A+C is bending away from us. This tail curves towards
the line that connects with members F and G and comprises the faint ionized
gas  around member E and the faint zone H. Member E could be an independent
kinematical entity. 

The slit positions at PA 133$^\circ$ and PA 128$^\circ$, indicate that the
brightest part of members F1 and F2 do not  show important velocity gradients.
PA 133$^\circ$, which was selected to cover precisely the main axis that
connects  both knots, shows a sinusoidal behavior although this is not seen in
the other position angle. In PA 128$^\circ$  (which does not cover exactly the
nucleus of F2) we see an apparent streaming motion towards positive radial 
velocities of the faint ionized gas at the north of F1. This streaming motion
seems to connect with the southern tip  of H. In any case, F1 and F2
seem to be  kinematically different to the tidal tail that emerges from the
southwest of the A+C complex. It is interesting to note  that the average radial
velocities of members E and F are quite similar, although they are relatively
distant objects  within the group. Their velocities are consistent with those
expected from their location in the \ion{H}{1} velocity  maps obtained by
\citet{W91}. Both objects are located in a zone where the velocity of the
neutral gas is  rather constant. 

Finally, member G was covered by PA 133$^\circ$. The morphology of this galaxy
indicates that it is a small disk  galaxy seen nearly face-on. It shows a
linear velocity gradient, with the radial velocity increasing towards the 
northwest. The amplitude of the velocity variation is of about 50 km s$^{-1}$.
This behavior indicates that this  galaxy is in solid-body rotation. The
velocity map of member G obtained by \citet{R03} shows a similar  velocity
gradient in this object; the radial velocity increases from the southeast to
the northwest.

\section{Discussion}

\subsection{Ages of the bursts and stellar populations}

The first determination of the age of the main burst of HCG~31 was performed by
\citet{R90}. Assuming that WR  stars are not contributing to the ionization of
the gas, they found that N(WR)/N(OB) $\sim$ 5. From this ratio, those  authors
suggested that this burst of star formation occurred 10 Myr ago and is rapidly
declining at the present.  IV97 determined the ages of the bursts making use of
$(U-B)$ versus $(V-I)$ star formation diagrams combined with the population 
synthesis models for instantaneous bursts described by \citet{LH95}. They found
an age between  3.16 and 10 Myr for the youngest bursts, but also detected an
old underlying stellar population in  some areas. IV97 also studied the
H$\alpha$  versus $U$ diagram detecting two bursts in
A+C and B: one about 10 Myr old, related to normal  star formation activity,
and other slightly younger due to star formation during the interaction. A
similar  behavior was also seen in member G. IV97 suggested that E is a young
member, but that the SFR has been decreasing in it for the last 10 Myr. For these
authors, member F shows an age almost coincident with the age of the  youngest
bursts in A+C and B; this object might have been created recently from gas
ejected during the interaction.

IV97 concluded that the most remarkable point is that the youngest episode in
all galaxies is almost simultaneous,  implying that it could have been
triggered by the interaction between galaxies A and C. In this way, they
suggested  that two episodes of star formation have occurred in HCG~31: the
youngest one seems to be the dominant in A+C and F;  the oldest one in E, and
the contribution of both episodes seems to be the same for B and G. 

H$\alpha$ emission traces recent star formation activity. W(\Ha) can be used to
estimate the ages of recent episodes  of star formation (e.g. \citealt{LH95}),
since it decreases with time. \citet{JC00} compared the observed  W(H$\alpha$)
with the predictions of \citet{LH95} models. They found that the W(\Ha) sources
are very  young (less than 10 Myr), with a peak in the distribution
corresponding to ages of about 5 Myr. A+C, B, and G show  signs of star
formation over the past 10 Myr, showing a peak in 5 Myr. They also reported
that B shows a second peak at about  10 Myrs, E shows young (1--3 Myr) star
formation and an older population similar to the other members, but F has only 
very strong W(\Ha), indicating an age below 4 Myr.

We have combined our broad-band optical and NIR photometric values with
STARBURST99 \citep{L99} models to estimate the age of the bursts. We have
chosen two different spectral synthesis models, both for an  instantaneous
burst with a Salpeter IMF, a total mass of 10$^6$ M$_\odot$ and a 100 M$_\odot$
upper mass, but with  two different metallicities: Z/Z$_\odot$ =0.4, and
Z/Z$_\odot$ =0.2. The metallicities we derive from the O abundance  in the
different objects (see Table~\ref{table5}) indicate that this metallicity range
should provide a good fit.


\begin{table*}[t]
  \caption{Age estimations of the bursts in HCG~31}
  \label{table6}  
  \scriptsize
  \begin{tabular}{ccccccccccc}
    \noalign{\smallskip}
    \tableline\tableline
	\noalign{\smallskip}
    Method               &  A    &   B   &  C    &   E   &   F1  &   F2  &   G   &  H    & Q \& D & Ref.\\
    \tableline
   H$\alpha$ vs. $U$              & 3--10& $\sim$10& 3--10 & $\sim$10  & 3--10&\nodata&$\sim $10&\nodata&\nodata 
&IV97\\
   $(V$ vs. $I)$+W(H$\alpha$)&$\sim5$& 5--10 &$\sim5$& 1--3 & $<$ 4&\nodata&$\sim $5&\nodata&\nodata  
&JC00\tablenotemark{d}\\
   \tableline
   \noalign{\smallskip}
    W(H$\beta$)\tablenotemark{a} & 6.5$\pm$1.5 & 7$\pm$1 & 5$\pm$1 & 5.5$\pm$1.5 & 2.5$\pm$0.5 & 2.5$\pm$0.5 & 7$\pm$1 
& 4.5$\pm$1.5 &\nodata& TW\tablenotemark{e}\\
	
    W[O III] \tablenotemark{a}  & 6.5$\pm$1.5 & 7$\pm$1 & 4.5$\pm$1.5 & 6$\pm$1 & 2.5$\pm$0.5 & 2.5$\pm$0.5 & 5$\pm$1 
& 4.5$\pm$1.5 &\nodata& TW\tablenotemark{e}\\
    W(H$\beta$)\tablenotemark{b} & 7$\pm$1 & 7$\pm$1 & 4.5$\pm$0.5 & 5.5$\pm$1.5 & 2.5$\pm$0.5 & 2.5$\pm$0.5 & 5$\pm$1 
& 4$\pm$1 &\nodata& TW\tablenotemark{e}\\
    $(U-B)$\tablenotemark{c}     & 7.5$\pm$3.5& 7.5$\pm$3.5& 8.5$\pm$3.5& 8$\pm$3& 3.5$\pm$0.5 & 3.5$\pm$0.5 & 
8.5$\pm$3.5&\nodata&$>$500 & TW\tablenotemark{e}\\
$(B-V)$vs.$(U-B)$\tablenotemark{c}&10.5$\pm$4.5& 10$\pm$5 & 10.5$\pm$4.5 & 10.5$\pm$4.5 & 4$\pm$1 & 4$\pm$1 & 
10.5$\pm$4.5 &\nodata&\nodata& TW\tablenotemark{e}\\
    $(V-R)$\tablenotemark{c}        & 6.5$\pm$1.5 & 7$\pm$1& 6$\pm$1& 7$\pm$1 & 3$\pm$1 & 3$\pm$1 & 7$\pm$1 & 
\nodata&$>$100 & TW\tablenotemark{e}\\
    $(V-J)$\tablenotemark{c}      & 9$\pm$2& 6.5$\pm$1.5 & 7$\pm$1 & 6.5$\pm$0.5 &   3$\pm$0.5   & 4.5$\pm$0.5 & 
7$\pm$1 &\nodata&$>$100 & TW\tablenotemark{e}\\
$(J-H)$vs.$(H-K_S)$\tablenotemark{c}&7.5$\pm$0.5 & 7$\pm$1 & 4.5$\pm$0.5 & 5.5$\pm$1.5 & 3.5$\pm$0.5 & 4.5$\pm$0.5 & 
7$\pm$1 &\nodata&\nodata& TW\tablenotemark{e}\\
   \tableline
   \noalign{\smallskip}
    Estimated age (Myr)    & 7$\pm$1 &   7$\pm$1  &  5$\pm$1   &    6$\pm$1 & 2.5$\pm$0.5    
              & 2.5$\pm$0.5   & 6$\pm$1 &    4$\pm$1     &$>$500 & TW\tablenotemark{e}\\
    \tableline\tableline
  \end{tabular}
  \tablenotetext{a}{\citet{SL96} models.}
  \tablenotetext{b}{\citet{SV98} WR galaxies models.}
  \tablenotetext{c}{STARBURST~99 \citep{L99} models.}
  \tablenotetext{d}{\citet{JC00} using HST data.}
  \tablenotetext{e}{This Work.}
  
\end{table*}

\begin{figure*}[t!]
\centering
\includegraphics[width=0.45\linewidth]{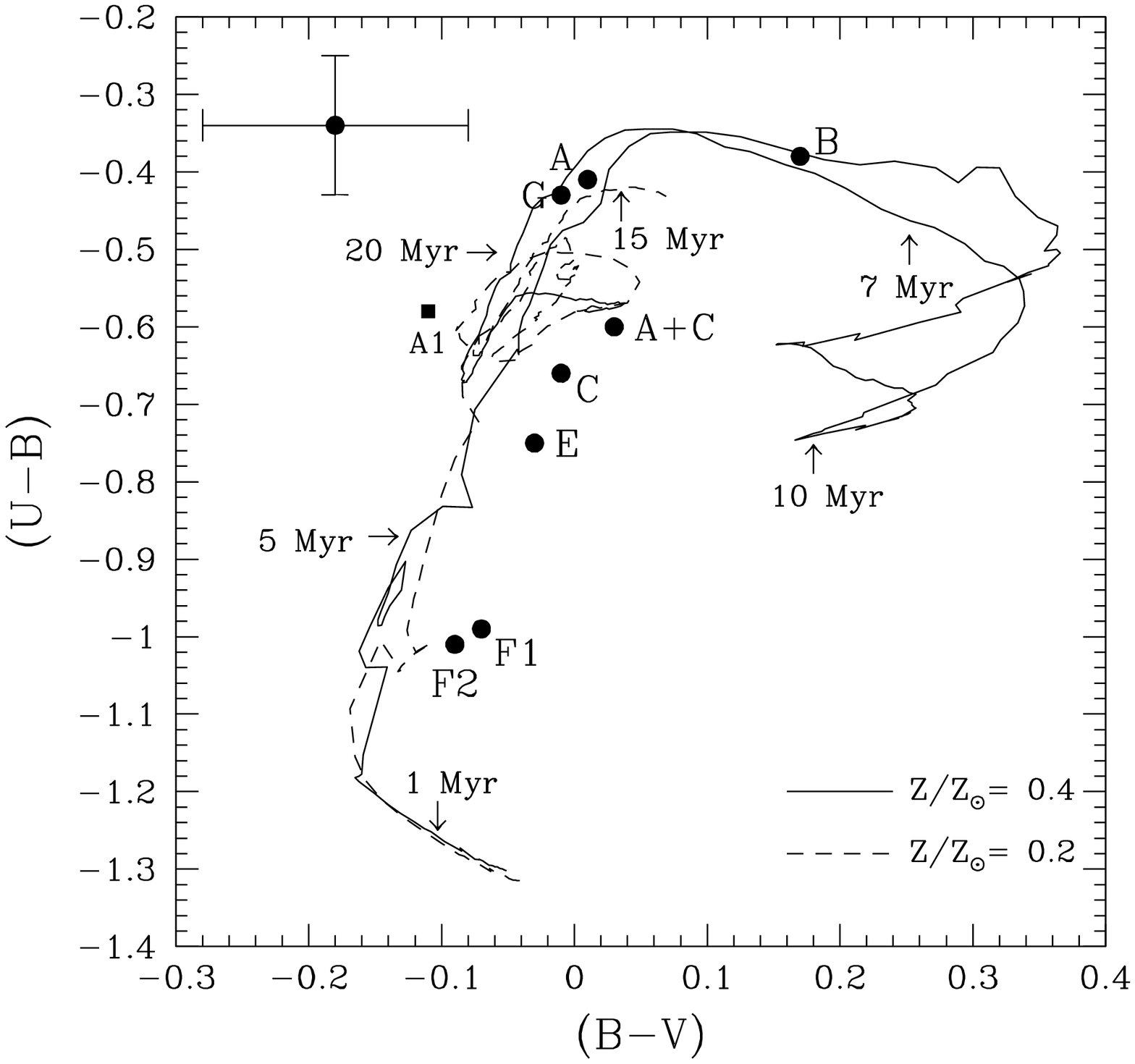}
\includegraphics[width=0.45\linewidth]{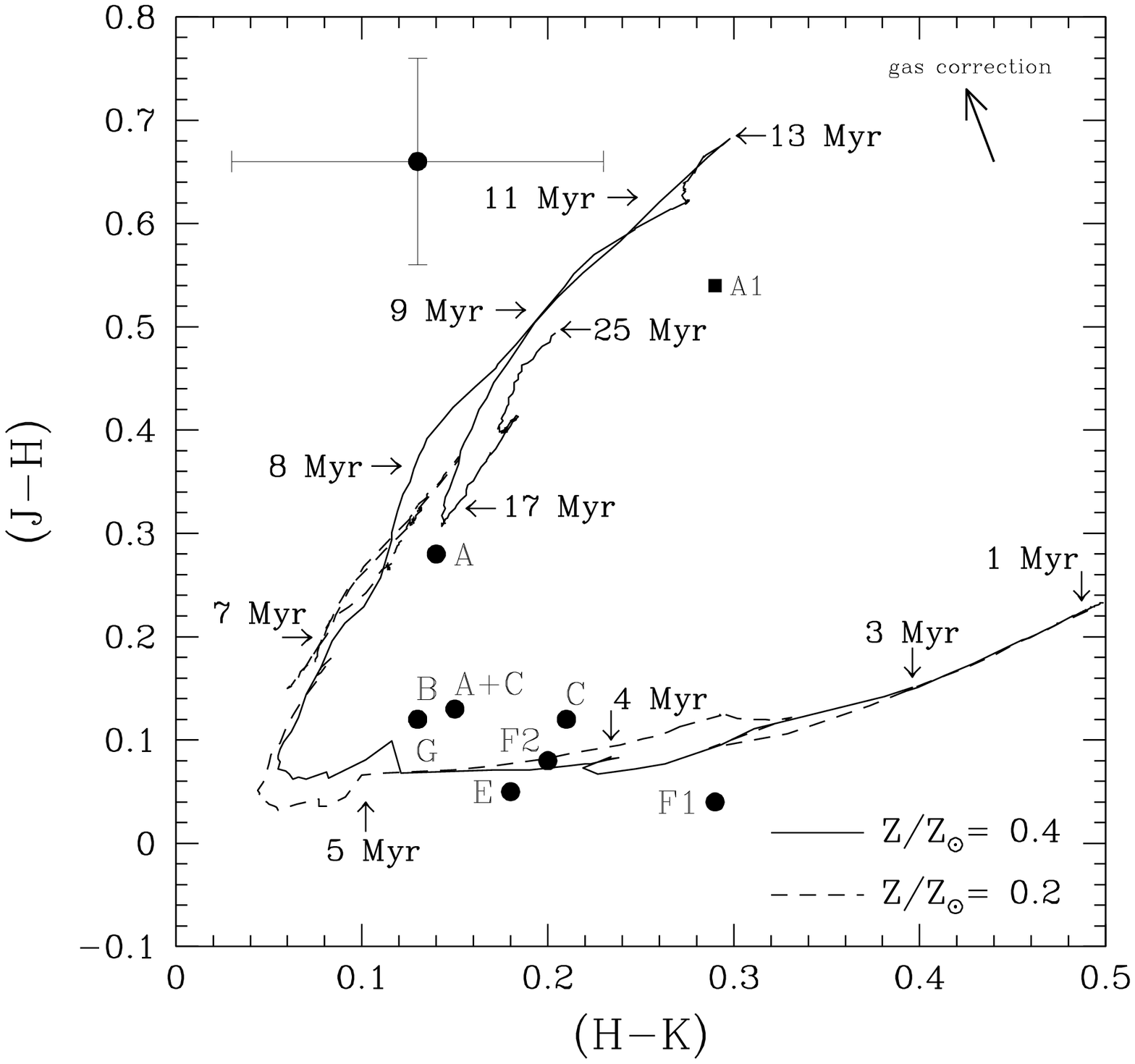}
\protect\caption[ ]{\small{(a) $(U-B)$ versus $(B-V)$ and (b) $(H-K_S)$ versus
$(J-H)$ values of the objects and STARBURST~99  \citep{L99} predictions for an
instantaneous burst with a Salpeter IMF. We have marked with arrows some ages
of the  $Z/Z_\odot$=0.4 model, showing the temporal evolution of the burst. The
error bars indicate the average uncertainties  in the data. We have included
the photometric values derived for A1 with a square.}}
\label{fig9}
\end{figure*}

Young bursts of star formation are expected to have very negative $U-B$ colors,
and most of the luminosity in both  bands should come from the massive star
populations. Therefore, $(U-B)$ can be used as a very convenient age estimator 
of the youngest stars. In this way, we have estimated the age of the different
galaxies and zones of HCG~31 comparing  their $U-B$ color and the results of
STARBURST~99 \citep{L99} selected models. The estimated ages are  shown in
Table~\ref{table6}. It should be remarked that with our spectroscopical data we
have obtained a direct  determination of  the interstellar reddening for most
of the galaxies of the group. This allows us to break the age-reddening 
degeneration in the broadband color-color diagrams (see \citealt{J99}). We have
found that members D and Q show  the highest $(U-B)$ values: they are basically
old galaxies, with an age $>$500 Myr, without evidences of intense ongoing 
star formation. On the other hand, the objects with the lowest $(U-B)$ colors
are F1 and F2. Therefore, they can be  interpreted as the youngest objects of
the group, with ages lower than 5 Myr. There is not a unique solution for the 
age of the rest of the objects due to the sinusoidal behavior of the $(U-B)$
color-age relation obtained from the  models. We can try to solve this
ambiguity making use of a color-color diagram plotting the observed values of 
$(U-B)$ versus $(B-V)$ and comparing with the theoretical models from 1 to 20
Myr (Fig.~\ref{fig9}a). We can note that  the  observational points fit the
models fairly well, providing (in principle) a good age determination. Again,
F1 and F2  seem undoubtedly to be the youngest objects. Unfortunately, we
cannot still disentangle the age ambiguity in the zone  with $(U-B)\leq-0.6$
and $(B-V)\geq-0.1$, the zone where most of the observational data lie. The
lack of  substantial departures in the position of the observational data with
respect to the theoretical positions for single  burst models indicate that the
contamination of the $(U-B)$ and $(B-V)$ colors by underlying older stellar 
populations is not relevant. In fact, this is directly reflected in the very
good agreement between the ages  estimated from the $(U-B)$ color alone and the
$(U-B)$ versus $(B-V)$ diagram. 


We have also used the NIR colors to obtain independent age estimations and to
explore the presence of underlying older  populations. It is well known that
NIR colors are very good indicators of the stellar population age once the
metallicity  is fixed (e.g. \citealt*{V00,V02}). Evolutionary synthesis models
(as those of STARBURST~99 and others) show that  the NIR colors of young
populations differ significantly from those of older ones. We have used the
$(V-J)$ color as an age  indicator and the results are given in
Table~\ref{table6}. It is remarkable that objects F1 and F2 give again the 
lowest values,  in agreement with the previous determinations. Again, Q and D
seem to be old systems, with colors indicating ages  larger than 100 Myr. In
Figure~\ref{fig9}b we plot the NIR colors $(J-H)$ versus $(H-K_S)$ of the
observed objects and the  results  of STARBURST~99 models for bursts at
different ages. The $(J-H)$ colors of the star-forming galaxies of HCG~31 are 
clearly those expected for young populations, since values of the order of
0.5-0.6 are the typical ones of evolved  populations \citep{V02}. To compare
the colors of the galaxies with the models it is necessary to  correct the
observations for ionized gas line emission, since STARBURST~99 only considers
the contribution of the nebular  continuum. We have estimated an average value
of the emission line contribution using the results for 24  starbursts by
\citet{C97} following the NIR study of 3 WR galaxies by \citet{V02}. An arrow
at the right-upper corner  of Figure~\ref{fig9}b shows this average
contribution. As in  the previous diagrams, the observational data are again in
quite a good agreement with the models, indicating the remarkably small
contribution of the old underlying stellar populations and the nebular line
emission. In fact, the ages  derived for each object from the $(J-H)$ versus
$(H-K_S)$ diagram are in very good agreement with those obtained from  optical
indicators. 

\begin{figure*}[h!]
\centering
\includegraphics[width=0.5\linewidth]{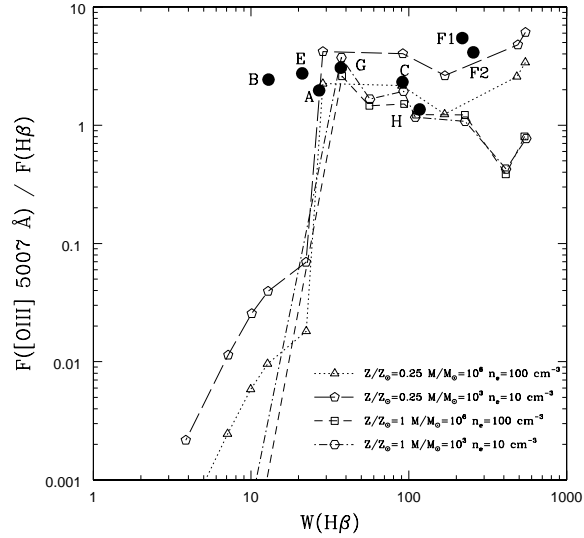}
\protect\caption[ ]{\small{F([\ion{O}{3}] $\lambda$5007) versus W (H$\beta$).
Models by \citet{SL96}. Tracks correspond  to sequences of different
metallicities and electronic densities. Each symbol marks the position of the
models at 1  Myr interval, starting in the upper-right corner of the diagram
with an age of 1 Myr. F1 and F2 are located between  the 2 and 3 Myr symbols.}}
\label{fig10}
\end{figure*}

Our spectroscopical data can be used to obtain additional independent age
indicators. We have used the models by  \citet{SL96} of \ion{H}{2} regions,
ionized by an evolving starburst embedded in a gas cloud of the same 
metallicity, to estimate the ages of the bursts. We have chosen two models,
both with metallicity Z/Z$_\odot$ = 0.25,  and with a total mass of M/M$_\odot$
= 10$^3$ and 10$^6$. In Figure~\ref{fig10} we plot our observational values of 
W(H$\beta$) and the [\ion{O}{3}] $\lambda$5007/H$\beta$ emission line flux (see
Tables~\ref{table3} and  \ref{table4}) and compare them  with the theoretical
models. All the objects show an excellent fit except member B, which has the
strongest  contribution of the underlying stellar continuum in the spectrum. As
for the rest of the indicators, objects F1 and F2  are the youngest bursts. In
Table~\ref{table6}, we also include the age estimations obtained from the
comparison of  the observed values of  W(H$\beta$) and W([\ion{O}{3}]) and the
results of the models by \citet{SL96}. The fact that both  indicators show
similar ages indicates that the underlying stellar continuum has a very limited
effect on the  measured W(H$\beta$). We have also used the models by
\citet{SV98}. In Table~\ref{table6}, we also include the age  estimations
obtained using these last models and the observed values of W(H$\beta$). As it
can be seen, the ages  obtained using both models are in excellent agreement. 



\begin{figure*}[h!]
\centering
\includegraphics[width=0.8\linewidth]{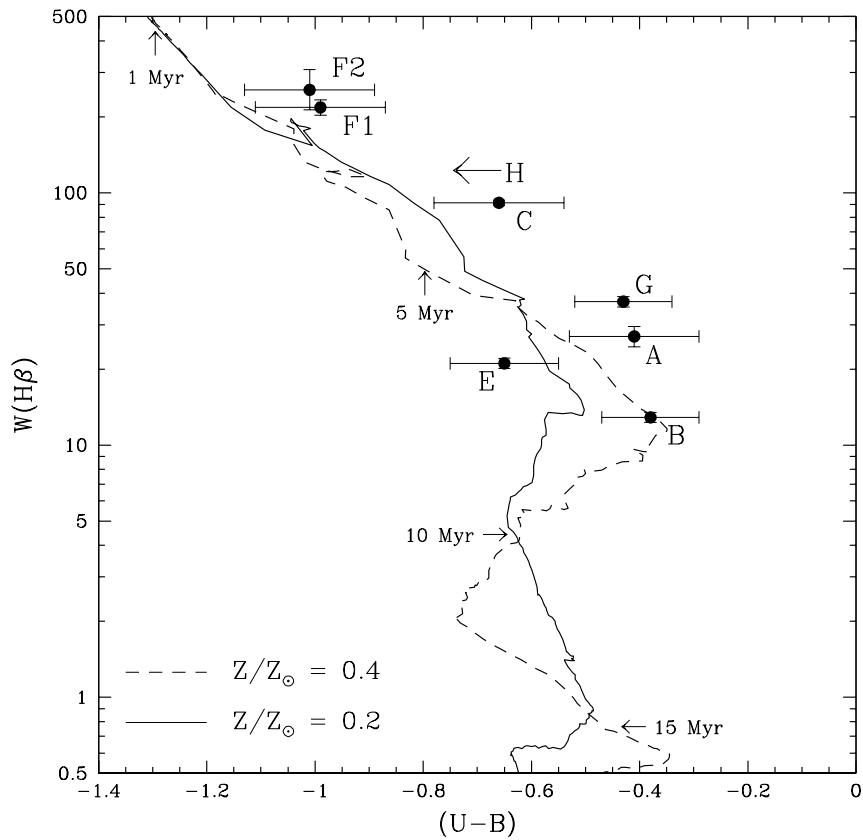}
\protect\caption[ ]{\small{W(\Hb) versus $(U-B)$ for members in HCG~31 and
STARBURST~99 \citep{L99} predictions for an instantaneous burst with a Salpeter
IMF. We have marked with arrows some ages of the $Z/Z_\odot$=0.4 model,
showing  the temporal evolution of the burst.}}
\label{fig11}
\end{figure*}

We have included our final adopted age estimation for each object in the last
row of Table~\ref{table6}. This  estimation is the  most probable 
value taking into account the indicators that are characteristic of the
emission from massive star populations: W(H$\beta$), W([\ion{O}{3}]), and
the $(U-B)$ color. We can see that, undoubtedly, objects F1 and F2  have the 
youngest bursts, with ages between 2 and 3 Myr. Member H also seems to be a
very young object, but the low  signal-to-noise ratio of its spectrum does not
permit us to be confident on this. Members A, B, E, and G show ages between 5 and
8 Myr in their bursts, whilst the brightest object of the group, member C,
shows an age between those  bursts and the younger F1 and F2. We think that it
is significant the consistency of the age estimations obtained  making use of
different theoretical models and very different indicators obtained from
independent kinds of data:  optical imaging, NIR imaging, and optical
spectroscopy. This indicates the remarkably small contribution of the 
underlying old stellar populations and the robustness of the actual knowledge
about starbursts and the models  available for their analysis. This can be
graphically demonstrated in the good agreement between photometric and 
spectroscopy data, as we show in Figure~\ref{fig11}, where we plot our
observational values of W(\Hb) versus $(U-B)$,  finding a very good agreement
with STARBURST~99 \citep{L99} predictions.  

\begin{figure*}[t!]
\includegraphics[width=0.43\linewidth]{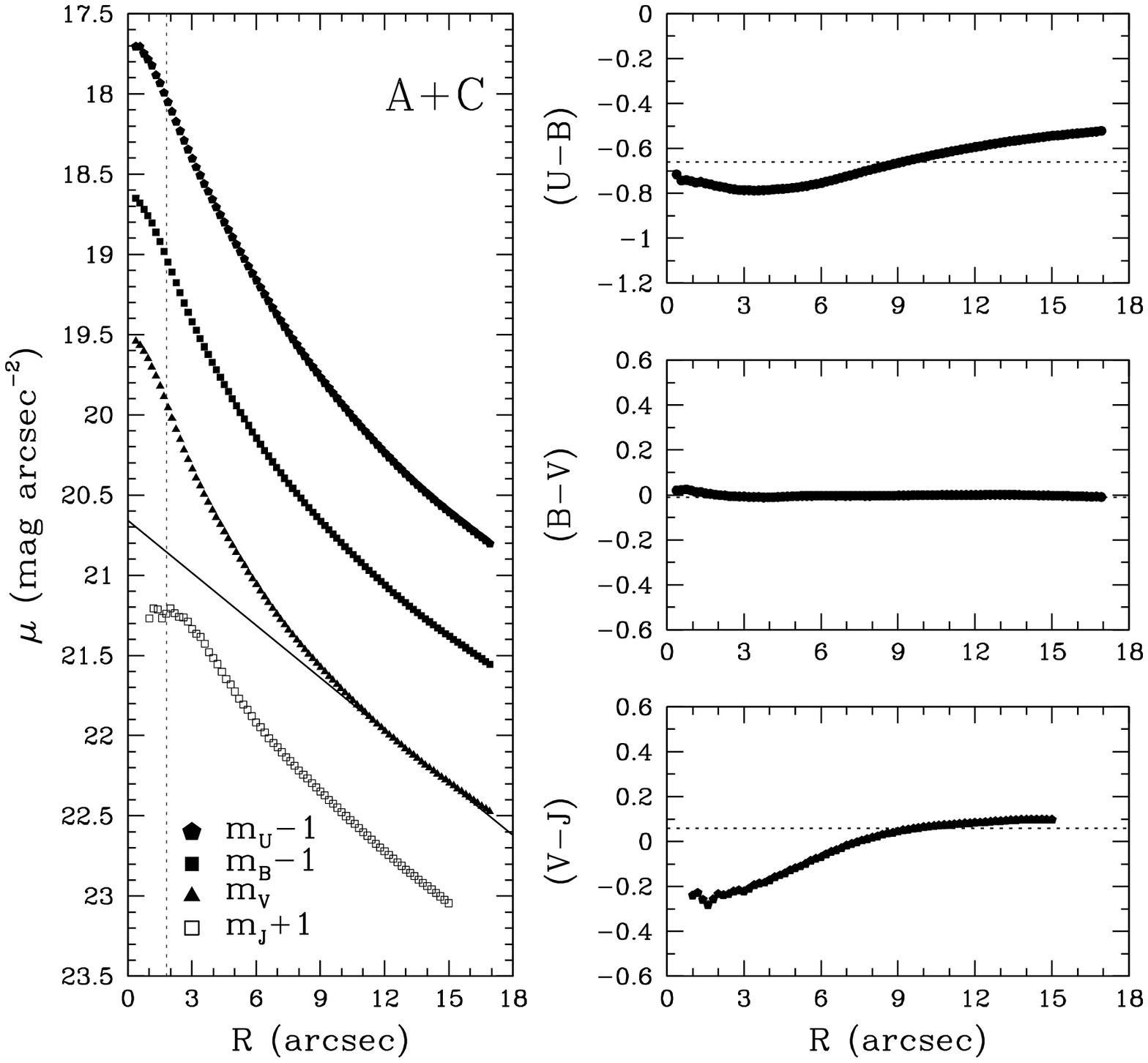}
\includegraphics[width=0.43\linewidth]{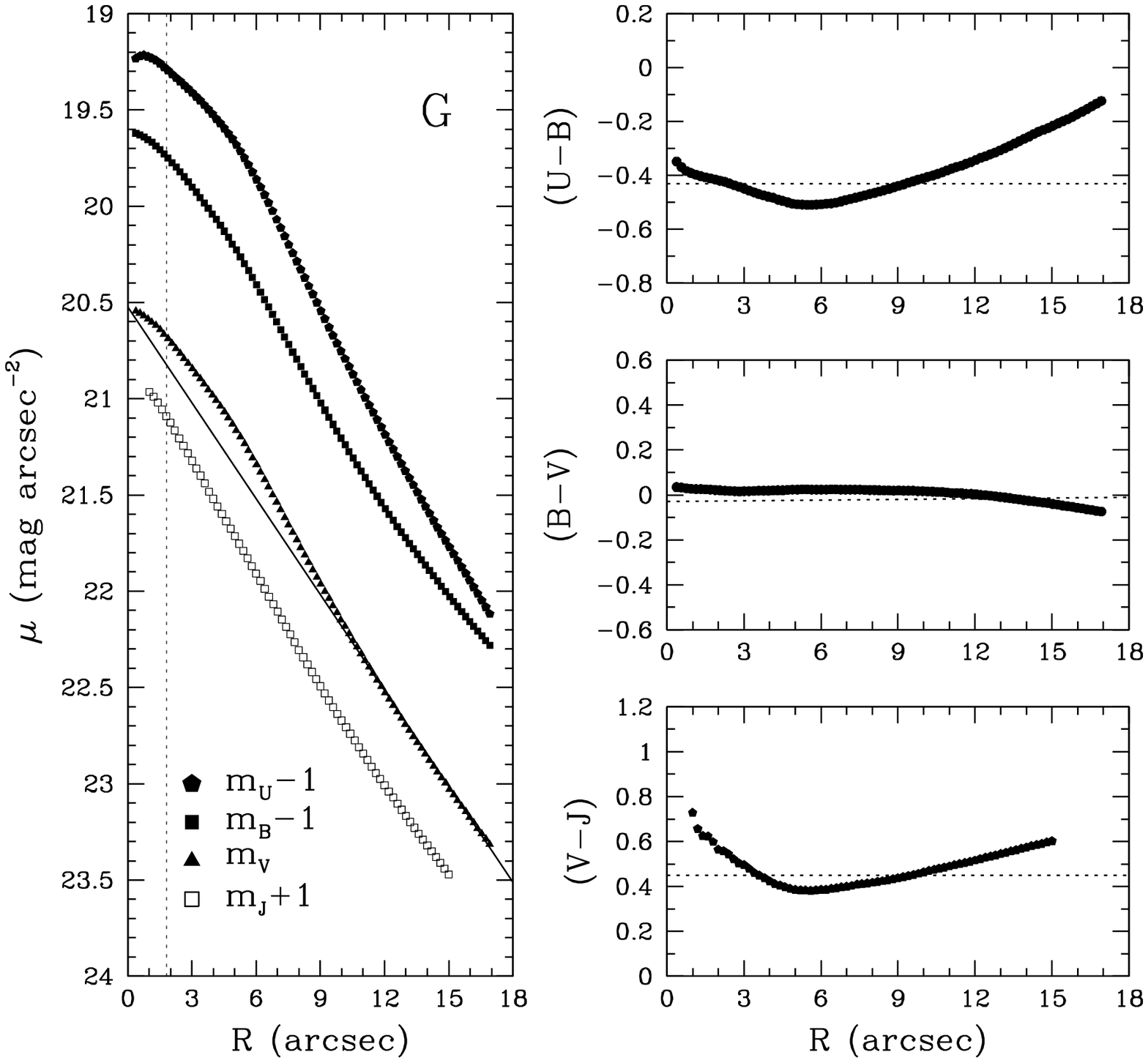}
\includegraphics[width=0.43\linewidth]{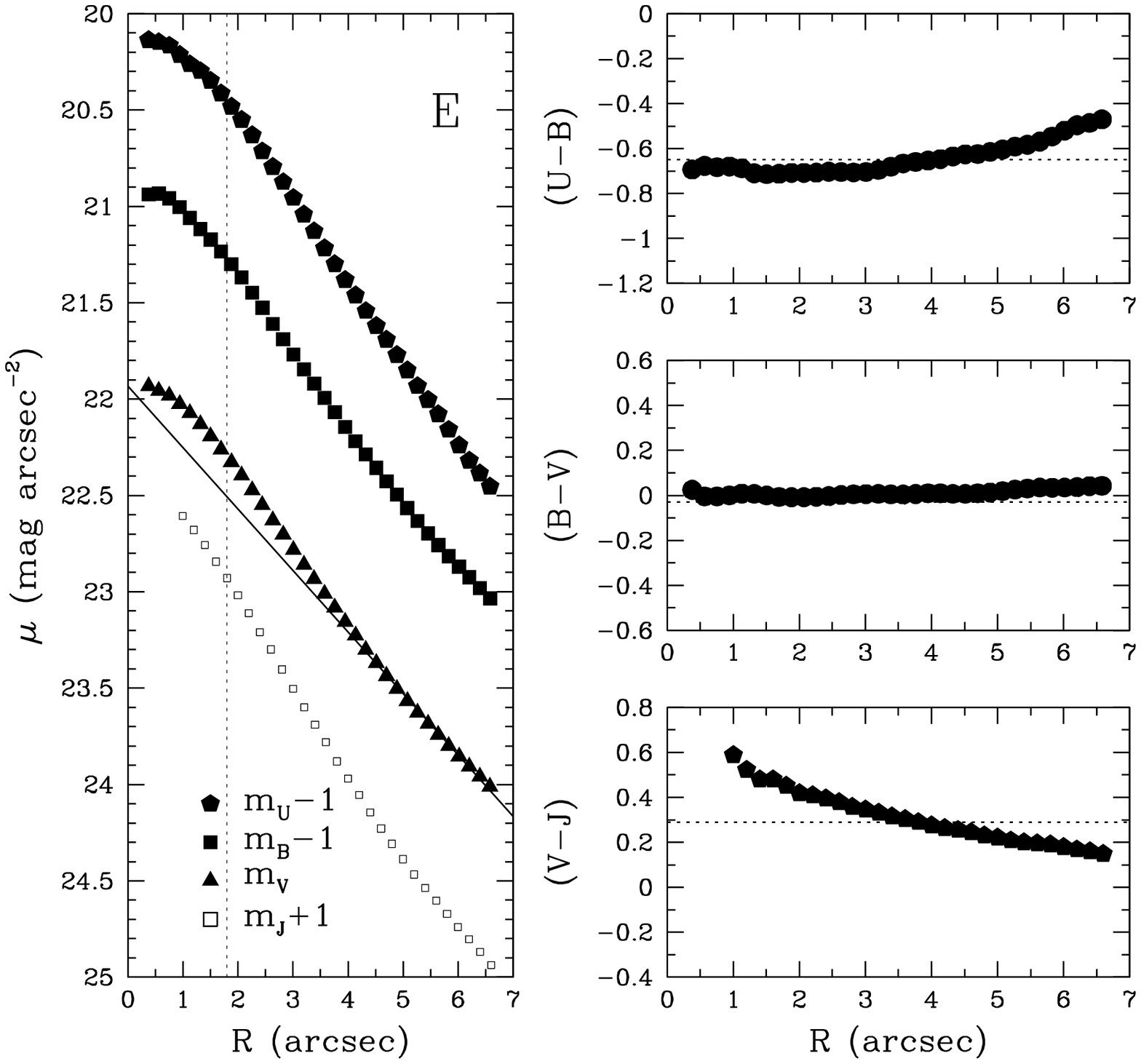}
\includegraphics[width=0.43\linewidth]{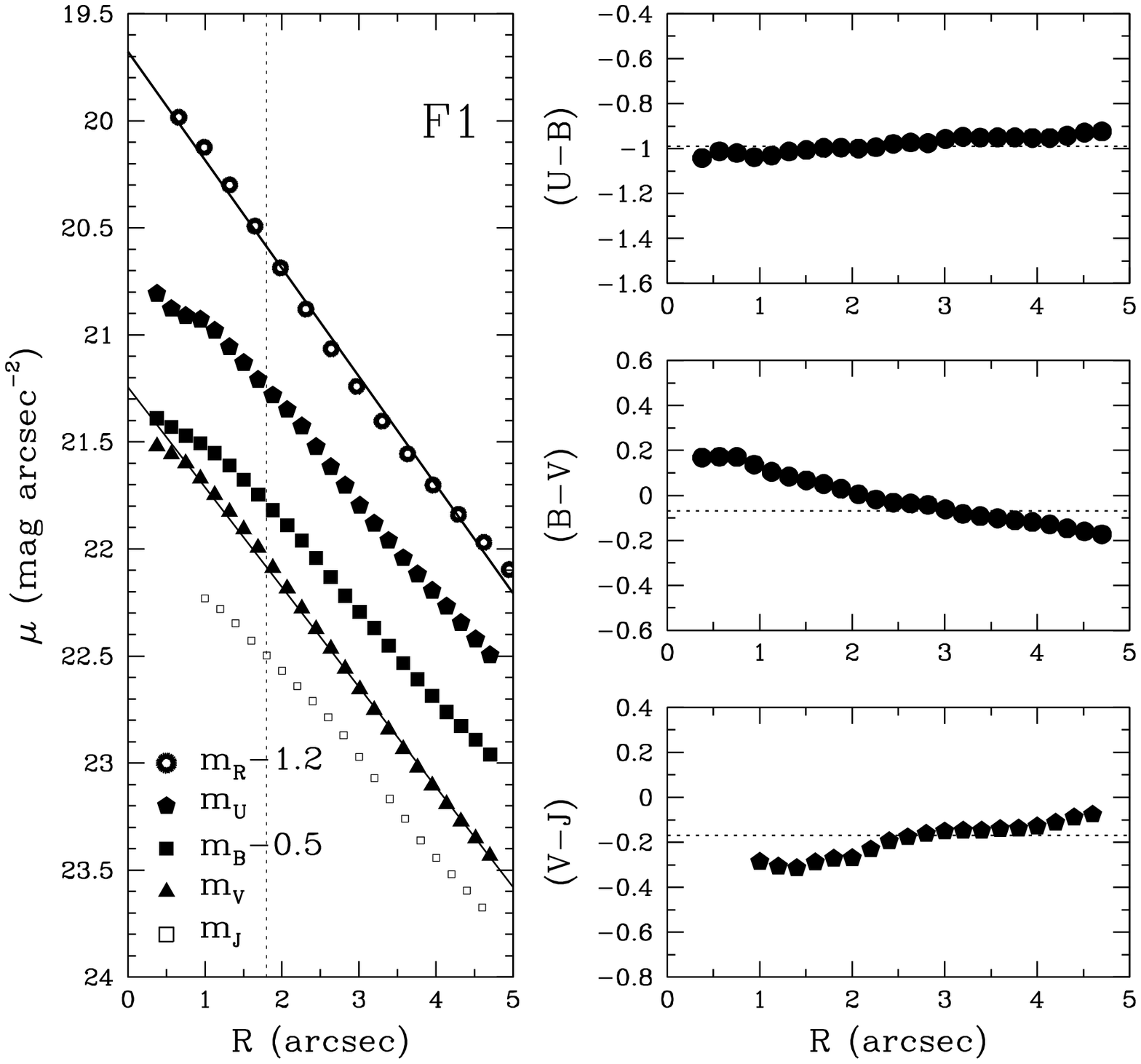}
\protect\caption[ ]{\small{Surface brightness and color profiles for the A+C complex and members
G, E, and F1 of HCG~31. The line in the  surface brightness diagrams is an
exponential law fitting to the $V$ profile, whereas the dotted vertical line is
the  average seeing.  The dotted horizontal line in the color profile diagrams
indicates the average color derived for  each system. For F1 we also show the
$R$ surface brightness profile and an exponential law fitting to it.}}
\label{fig12}
\end{figure*}

We have analyzed the surface brightness profiles of members A+C, E, F1, F2, and G
in order to study the spatial  distribution of their different  stellar
populations and to look for an underlying old component. We have taken
concentric surfaces at different radii  from the centre of each system, and
calculated the integrated flux inside each circle of area $A$ using the
relation:
\begin{eqnarray}
\mu_X=m_X+2.5 \log A ,
\end{eqnarray}
to obtain the surface brightness, $\mu$ (in units of mag arcseg$^2$), of each
circle $A$ (in units of arcseg$^2$) in  each filter $X$ ($U$, $B$, $V$ and
$J$), $m_X$ is the magnitude in the filter $X$. In Figure~\ref{fig12} we
present the  surface  brightness for $U$, $B$, $V$ and $J$ filters versus the
size of the aperture (in arcsec) for A+C, G, E, and F1. The dotted vertical  line
is the average seeing. This analysis is not precise for the study of irregular
morphologies like those of B and A+C, but it  is useful for our main purpose and
it is valid for approximately circular compact objects like E, F1, and F2. A 
detailed study on surface brightness profiles can be found in \citet{CV01}.
These authors  perform such  a study for a sample of blue compact dwarf
galaxies, which included several members of HCG~31: Mrk 1089 (A+C complex)  and
Mrk 1090 (member G). \citeauthor{CV01} find that the A+C complex presents an
exponential profile, whereas member G shows a  reasonable fit to both
exponential and de Vaucouleurs profiles. In the majority of the observed
galaxies,  \citeauthor{CV01} find a low surface brightness component underlying
the contribution of the starburst, but they do not  observe this in the case of
the galaxies of HCG~31. 



\begin{table}[t]
  \caption{Structural parameters of some galaxies in HCG~31. }
  \label{table7}  
  \small
  \begin{tabular}{cccc}
   \noalign{\smallskip}
    \tableline\tableline
	\noalign{\smallskip}
   Member &      U.C.\tablenotemark{a} & $\mu_{V,0}$& $\alpha_V$ (kpc) \\
    \tableline
	\noalign{\smallskip}
 	A+C  & yes&  21.66     &  2.38	  \\
	E  &   ?&  22.12     &  0.82	  \\
	F1 &  no&  22.25     &  0.56	   \\
	F2 &  no&  23.43     &  2.52	   \\
	G  &   ?&  21.52     &  1.58	   \\
   \tableline\tableline
  \end{tabular}  

  \tablenotetext{a}{Underlying component.}

 \end{table}


In Figure~\ref{fig12} we also show the radial color profiles $(U-B)$, $(B-V)$
and $(V-J)$ derived by a direct subtraction of  two light profiles. The color
profiles cannot be used to detect small variations in the inner regions of
the  galaxy, but they are useful for describing the color variation as a
function of the distance. In these graphics we  also indicate the
representative average color derived for each system (see Table~\ref{table2})
by a dotted  horizontal line.  Thus, this line shows the color within the
3$\sigma$ radio that we used for the aperture photometry. We  have centered the
radial profile in A+C just between the two brilliant central knots (see
Fig.~\ref{fig1}). In  Figure~\ref{fig12} we can see that A+C has a low surface
brightness component underlying the burst since the surface brightness profile
can be separated in two components, a disk structure and the contribution of 
the starburst. This underlying component was not detected by \citet{CV01}, but
it can also be observed in the $(U-B)$ and  $(V-J)$ color profiles. We have
performed an exponential law fitting to the $V$ profile, following the
expression:
\begin{eqnarray}
I=I_o \exp(-\alpha \, r),
\end{eqnarray}
that describes a typical disk structure: $I_0$ is the central intensity and
$\alpha$ is the scale length. The  fitting structural parameters are indicated
in Table~\ref{table7} and the fit is plotted over each $V$ profile in
Figure~\ref{fig12} with a straight  line. In member G the fitting is fairly
good, and this suggests that G has not an important underlying old component. 
However, the color profiles, specially $(V-J)$, show a strange behavior, that
can be explained noticing that the starbursts  in G are not located at the
center of the galaxy, but in its northwest side (see Fig.~\ref{fig1}). E could possess
a very  faint  underlying component, but we cannot confirm it. The exponential
fitting for this member was only performed for radii larger  than the one
considered for the optical photometry. Finally, we do not detect low brightness
underlying components in F1 and F2 in the optical filters (we have also included in 
Figure~\ref{fig12} the $R$ profile for F1 and its linear fitting). It can be seen 
that the old stellar population is not significant in these objects, which are 
clearly dominated by the starburst. This fact is in very good agreement with
our age estimation for them (see Table~\ref{table6}). \citet{JC00} also 
remarked that F1 and F2 were the only members in HCG~31 that did not show old populations.
However, following the recent study by \citet{N03},
a probable old population could be found if F1 and F2 were studied using high spatial
resolution photometry in NIR, because the starburst emission overshines the underlying 
population in the inner part of the objects in the optical wavelengths. Nevertheless, 
we argue that F1 and F2 seem to be \emph{really} very young and recently-formed objects, 
and that the relative contribution of an hypothetical old population, if any,
should be very small.

\subsection{WR population}

We have used evolutionary synthesis models by \citet{SV98} for O and WR
populations in young starburst to  estimate the number of O and WR stars in F1
and F2. We have assumed that L(\Hb)=4.76$\times 10^{36}$ erg s$^{-1}$ for a  O7V
star and L(WNL 4686)=1.7$\times 10^{36}$ erg s$^{-1}$ for a WNL star
\citep{VC92}. The extinction-corrected flux of  the broad \ion{He}{2}
$\lambda$4686 emission line is 4.69$\times 10^{-17}$ erg cm$^{-2}$ s$^{-1}$ for
F1 and  2.26$\times 10^{-17}$ erg cm$^{-2}$ s$^{-1}$ for F2 (see 
Table~\ref{table8}). However, we must correct these values for the size of the
slit with respect to the true area of each member. From the photometric values,
we have adopted a ratio of 4 for F1 and 2 for  F2. Taking into account a
distance of 54.8 Mpc for HCG~31, we obtain a luminosity for the \ion{He}{2}
$\lambda$4686  line of about 4.27$\times 10^{39}$ erg s$^{-1}$ for F1 and
1.71$\times 10^{39}$ erg s$^{-1}$ for F2. If we assume that all  the contribution
in the \ion{He}{2} $\lambda$4686 emission line comes from WNL stars, we obtain
around 40 and 10 WR  stars in F1 and F2, respectively. From the H$\beta$ flux,
correcting the contribution of the WR stars to the ionizing  flux, and the
contribution from other O stars (we have derived that $\eta\equiv$O7V/O=0.25 for
C and $\eta$=0.5 for F1  and F2 using the estimated age derived for each burst,
see Figure 21 in \citealt{SV98}), we finally obtain about 1660  and 680 O stars
and a WR/(WR+O) ratio of 0.024 and 0.014 for F1 and F2, respectively.

We have also performed a similar analysis for C. We obtain a WR/(WR+O) ratio of
0.003. This value is similar to the  one of 0.005 estimated by \citet{GIT00}. 

\begin{figure*}[t!]
\centering
\includegraphics[width=0.45\linewidth]{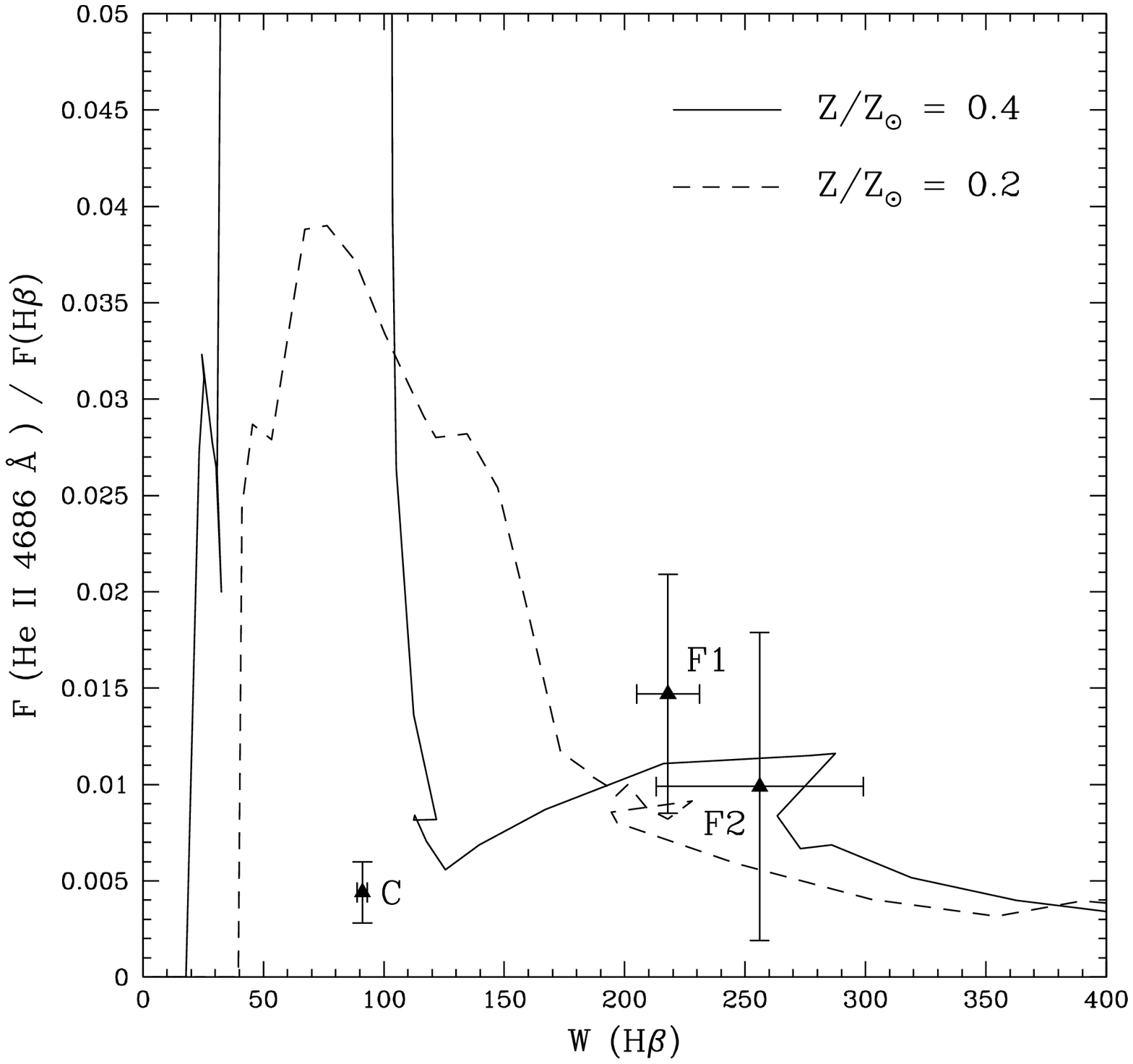}
\includegraphics[width=0.45\linewidth]{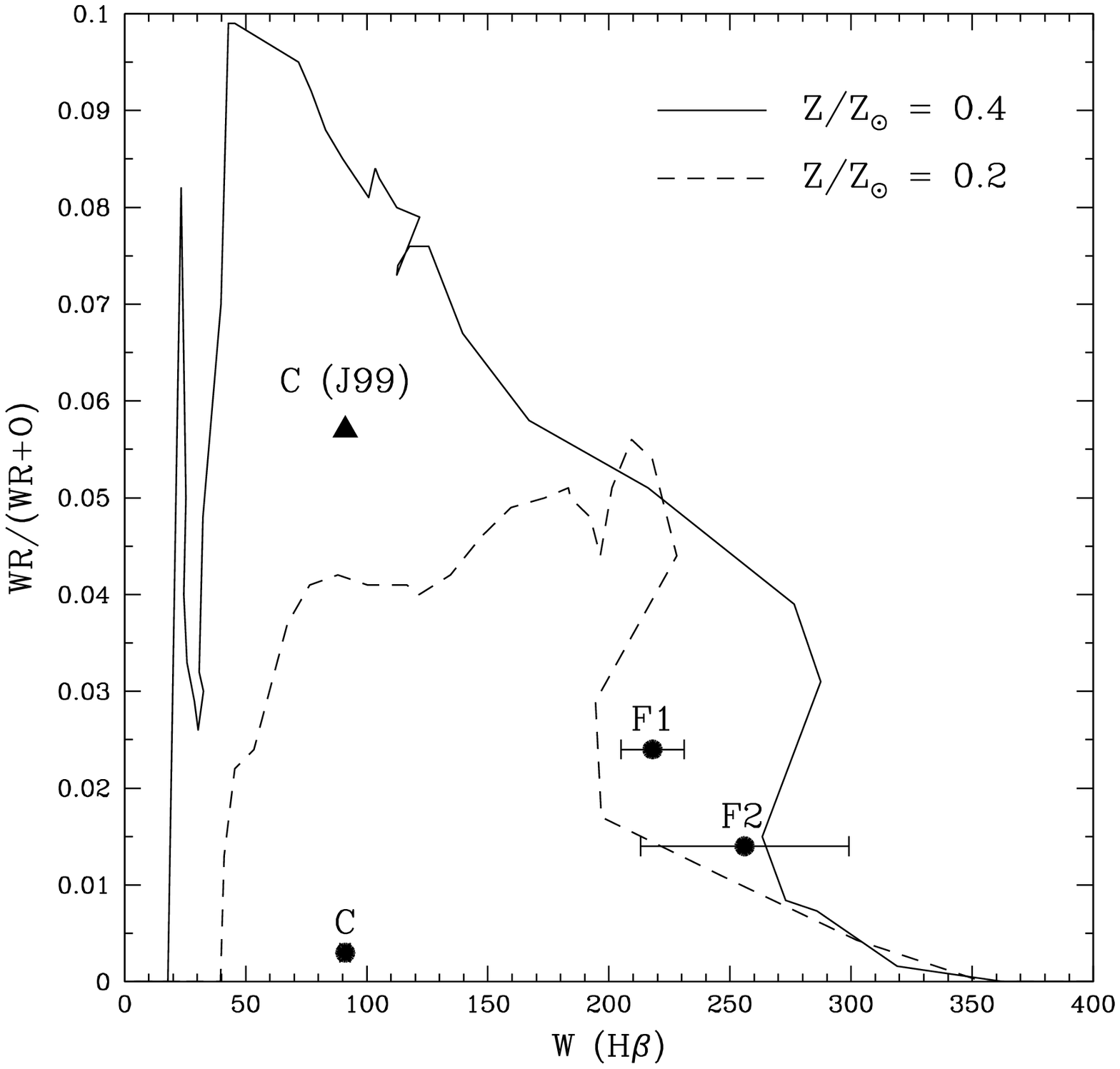}
\protect\caption[ ]{\small{(a) F(\ion{He}{2}) versus W(\Hb) for \citet{SV98}
starbursts models compared with our  results for C, F1, and F2. (b) WR/(WR+O)
versus W(\Hb) for \citet{SV98} models. We include our results from  optical
espectroscopy (\emph{circles}) and the one obtained from the \ion{He}{2}
$\lambda$1640 emission line flux  (\emph{triangle}) by \citet{J99} using
\citet{SV98} calibration (their eq. 18).}}
\label{fig13}
\end{figure*}

We show all the observed and derived WR population values for C, F1 and F2 in
Table~\ref{table8}. We have compared  these results with the predictions from
\citet{SV98} models for O and WR populations in young starbursts.  In
Figure~\ref{fig13}a we plot the \ion{He}{2} $\lambda$4686 emission line flux
versus the \Hb\ equivalent width for the  models with 0.4 and 0.2 $Z_\odot$. We
observe than F1 and F2 show quite a good agreement with these models, but C
does  not. This difference between the observed and predicted values in C seems
to be real, because \citet{GIT00} obtain a  similar WR/(WR+O) ratio. The
difference could be due to aperture effects. Perhaps, the area observed inside
the A+C  complex does not comprise the WR-rich area. This effect is not
important in F1 and F2 because of their small size.  Very probably, we are
measuring all the emission from the WR stars located in these bursts. 

In Table~\ref{table8} we also include the WR/(WR+O) ratio obtained using two
different calibrations by \citet{SV98}:  one from the blue WR bump flux (and their
eq. 17) and other derived, only for C, from the \ion{He}{2} $\lambda$1640 emission 
line equivalent width (and their eq. 18). We have assumed that all the blue WR bump
flux comes from the \ion{He}{2}  $\lambda$4686 emission line. We find a good
agreement for F1 and F2, but not for C. We have also used the  W(\ion{He}{2})
$\lambda$1640 obtained from HST UV spectroscopy by \citet{J99} for C, and the
calibration derived by  \citet{SV98} for this particular emission line (their
eq. 18), and find WR/(WR+O)$\approx$0.057, one order of  magnitude higher
than our estimations using the blue bump flux and the one by \citet{GIT00}.
This value, that is in  better agreement with the theoretical predictions, is
plotted with a triangle in Figure~\ref{fig13}b. The difference  between the UV
and optical values can be due to the fact that the UV spectrum was obtained from
HST observations of the bright knots of the A+C complex,
precisely the regions where the WR stars and the highest SFRs  are found.
Therefore, these determinations based on the W(\ion{He}{2}) $\lambda$1640
should be less affected by  aperture effects.


\begin{table*}[t]
\caption{Analysis of O and WR populations in C, F1, and F2.} 
  \small
  \label{table8}  
  \smallskip
  \begin{tabular}{lccc}
    \tableline\tableline
	\noalign{\smallskip}
                   &  C              &    F1           &  F2       \\
	\noalign{\smallskip}
    \tableline
    \noalign{\smallskip}
$F$(\ion{He}{2} $\lambda$4686)\tablenotemark{a}&  2.04$\times10^{-16}$&  4.69$\times10^{-17}$ & 2.26$\times10^{-17}$   
\\
     W(\ion{He}{2} $\lambda$4686) (\AA)       &  0.34$\pm$0.07      &  2.66$\pm$0.5        & 2.03$\pm$0.6  \\	
   Slit size correction &    7.3          &  4              & 2   \\
   Size (kpc$^2$)       &     6.43        & 0.98            & 0.49 \\
   $\eta$               &     0.25        & 0.50            & 0.50 \\
   WNL stars            &   300           &  40             & 10  \\
   O stars              & 96600           & 1660            & 680 \\
   WR/(WR+O)            & 0.003           & 0.024           & 0.014 \\
   WR/(WR+O)\tablenotemark{b}     & 0.008           & 0.022           & 0.015 \\
   WR/(WR+O)\tablenotemark{c}     & 0.057           & \nodata         & \nodata \\
   \noalign{\smallskip}
    \tableline\tableline 

  \end{tabular}
   \tablenotetext{a}{In units of erg s$\rm^{-1}$ cm$\rm^{-2}$.} 
   \tablenotetext{b}{Based on our determinations of the blue WR bump flux.}    
   \tablenotetext{c}{Based on the W(\ion{He}{2} $\lambda$1640) obtained by \citet{J99}.}
\end{table*}


\subsection{Star formation rates}

We can use the \Ha\ emission to calculate the SFR in each burst of HCG~31. 
IV97 presented the \Ha\ luminosities obtained from CCD imagery for all the
members in HCG~31. We have used their data  and the \citet{K98} calibration:
\begin{eqnarray}
SFR_{H\alpha} = 7.94 \times 10^{-42}L(H\alpha),
\end{eqnarray}
to estimate the SFR (in units of \Mo\ yr$^{-1}$) in the most important members
from L(H$\alpha$) (in units of erg  s$^{-1}$). But we have to correct their
data from several contributions.  First, we correct by the different
distance assumed for HCG~31 (IV97 considered that the compact group was at
41.23  Mpc whereas we are using a distance of 54.8 Mpc). Then, we must
also correct the data for the reddening that we  have derived for each burst.
Finally, we correct  for the contribution of [\ion{N}{2}] emission to the total
\Ha\ flux taking into account our spectroscopic data. That  contribution
represents between 5 and 15 \% of the narrow \Ha\ filter total flux. We show
the corrected \Ha\  luminosity and the SFRs derived from them in
Table~\ref{table9}.


\begin{table*}[t]
\caption{Star formation rates for HCG~31 bursts.}
  \label{table9} 
  \footnotesize
  \smallskip
  \begin{tabular}{ccccc}
    \tableline\tableline
	\noalign{\smallskip}
   Member &  log L(\Ha) & SFR (\Mo\ yr$^{-1}$) &  SFR (\Mo\ yr$^{-1}$ kpc$^{-1}$)\tablenotemark{a} &  
Reference\tablenotemark{b}     \\
	\noalign{\smallskip}
    \tableline
    \noalign{\smallskip}
	A+C  &  41.78  &  4.77  & \nodata &  IV97 adapted \\
	A+C inner part  &  41.54  &  2.74  & 0.43    &  Spec, $r=7.3$, $A=6.43$ kpc$^{2}$ \\
	A1   &  38.76  &  0.005 & 0.01    &  Spec, $r=2$, $A=0.42$ kpc$^{2}$ \\
	B    &  40.22  &  0.13  & 0.04    &  Spec, $r=6$, $A=2.94$ kpc$^{2}$\\
	E    &  40.50  &  0.27  & \nodata &  IV97 adapted \\
	E    &  40.05  &  0.09  & 0.05    &  Spec, $r=7$, $A=1.715$ kpc$^{2}$\\ 
	F1   &  40.40  &  0.20  & \nodata &  IV97 adapted \\
	F1   &  40.12  &  0.10  & 0.10    &  Spec, $r=4$, $A=0.98$ kpc$^{2}$\\
	F2   &  40.09  &  0.10  & \nodata &  IV97 adapted \\
	F2   &  39.67  &  0.04  & 0.08    &  Spec, $r=2$, $A=0.49$ kpc$^{2}$\\
	G    &  40.75  &  0.41  & \nodata &  IV97 adapted \\
	G    &  40.60  &  0.31  & 0.06    &  Spec, $r=10$, $A=4.9$ kpc$^{2}$\\
	H    &  38.17  &  0.001 & 0.015   &  Spec, $r=1$, $A=0.26$ kpc$^{2}$\\
   \noalign{\smallskip}
    \tableline\tableline
	
  \end{tabular}
  \tablenotetext{a}{Derived considering the SFR shown in column 3 and the total area $A$ of the burst shown in column 
5.}
  \tablenotetext{b}{The $r$ parameter shows the ratio between the total area of the burst and the area covered by the 
slit.}
\end{table*}


We have also estimated the SFR from the \Ha\ luminosity provided by our spectra
for comparison. We have  considered the slit size with respect to the whole
area in each burst to estimate the total luminosity ($r$ factor  in
Table~\ref{table9}). We include the results obtained by this method in
Table~\ref{table9}, as well as the SFR per kpc, that was calculated dividing 
the SFR by the total area, $A$, of the observed burst. We note that there is a good
general agreement between both methods, although the results derived from the
IV97 data are systematically higher by a factor of 2. As we could expect, the
highest SFR is found at the  center of the A+C complex, with 4.77 \Mo\
yr$^{-1}$ (from the IV97 data) and 2.74 \Mo\ yr$^{-1}$ (the spectroscopic
value). We can  compare these values with the SFR derived from the 60 and 100
$\mu$m fluxes measured with \emph{IRAS}  \citep{Mo90}. \citeauthor{Mo90}
found $f_{60}$=3.920 Jy and $f_{100}$=5.840 Jy. Using the \citet{K98}
calibration:
\begin{eqnarray}
SFR_{IR}= 4.5 \times 10^{-44} L_{FIR},
\end{eqnarray}
where
\begin{eqnarray}
L_{FIR} = 1.26 \times 10^{-11} \big( 2.58 f_{60} + f_{100} \big),
\end{eqnarray}
we find  that the SFR derived from infrared fluxes is $SFR_{IR}$=3.18 \Mo\
yr$^{-1}$, in good agreement with our  estimations from \Ha\ luminosities. We
can also remark that C, F1, and F2 are the systems with the highest SFRs when the 
size of each burst is considered.

\subsection{The luminosity-metallicity relation}

\begin{figure*}[h!]
\centering
\includegraphics[width=0.5\linewidth]{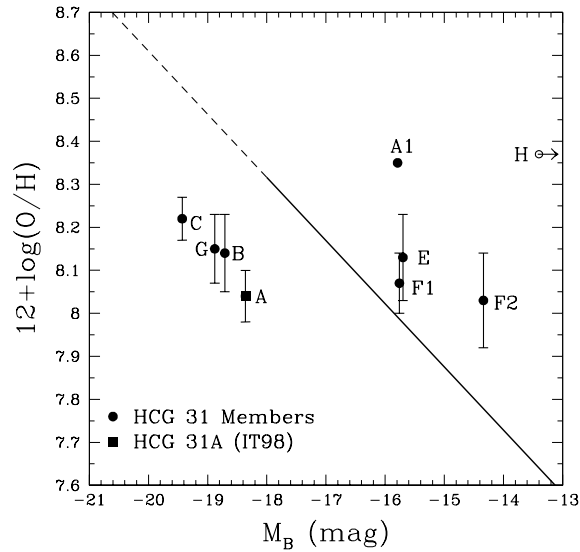}
\protect\caption[ ]{\small{The relation between luminosity and metallicity for
the galaxies in HCG~31. The solid line  is the relation for dwarf irregulars
found by \citet{RM95}, while the dashed line is an extrapolation of  it. The
O/H ratios for A1 and H are estimations based on empirical calibrations. The
O/H ratio for member A is from \citet{IT98} (IT98).}}
\label{fig14}
\end{figure*}

\citet{R03} used the relation between absolute magnitude and metallicity for
dwarf irregular galaxies proposed by  \citet{RM95} for the HCG~31 members,
finding that the position of A and C is rather unusual.  In fact, these objects
seem to be too luminous for their oxygen abundances. In Figure~\ref{fig14}, we
show the  luminosity-metallicity relation we obtain for the objects as well as
the \citet{RM95} relation for  comparison (as in \citealt{R03}, with an
extrapolation to higher luminosities). Our O/H ratios are quite  similar to
those adopted by \citet{R03} for members C and F. In the case of member B our
determination is  0.2 dex lower than that obtained by \citet{R03} and now is
similar to the abundances of the rest of the  brightest objects of the group.
In the case of member A, we only have an indirect estimation of the O/H ratio
of A1  based on empirical methods. Taking into account the systematic larger
abundances that we obtain with the empirical  methods for all the objects (see
\S~3.3.2), we prefer to use the direct determination of O/H obtained by 
\citet{IT98} as representative for A. With this abundance, the position of A is
similar to the other objects on the  diagram. The position of object H in the
diagram is very uncertain because we only have a lower limit for its absolute 
magnitude and the O/H ratio has also been estimated from empirical
calibrations. 

In Figure~\ref{fig14}, we can see that the unusual position  of A and
C in the luminosity-metallicity relation found previously by \citet{R03} is
also a property of the other bright galaxies of the group: B and  G. Therefore,
members A, B, C, and G (the four brightest galaxies of the HCG~31 group) show
rather similar O/H ratios  (12+log(O/H) = 8.0--8.2), about 0.3 dex
(a factor of 2) lower than the value expected from the \citet{RM95}  relation.
We consider that this is a real behavior because the abundances are well
determined, based on the direct measurement of the electron temperature for most
of the objects. In addition, as we have commented above, our  abundances are
similar to previous determinations for those objects with data available in the
literature. The  systematical offset of the objects in the
luminosity-metallicity relation can also be explained by a displacement of 
about 2 magnitudes in the magnitude axis. This amount seems too large to be
accounted for by observational errors.  Our absolute magnitudes are similar to
those obtained in previous works (see \citealt{R03}) and the colors  derived
(even comparing optical and NIR data) do not show any systematic trend that
could suggest the presence of  calibration problems.


The apparently unusual position of the brightest galaxies of HCG~31 in the
luminosity-metallicity relation deserves  further discussion. \citet{R03}
interpreted the unusual positions they found for galaxies A and C as the 
combination of dilution due to strong gas inflow (arising from the funneling of
the gas towards the central zones due to interactions) and the transient high
luminosity due to the strong star formation. Taking into account our results, 
the explanation proposed by \citet{R03} based on inflows seems unlikely. It
is difficult to understand  how the four galaxies could experience gas inflows
in a manner that produce similar final abundances in all the  objects. The four
galaxies are located in different positions inside the group and have different
physical  characteristics; therefore, the properties and intensities of the
hypothetical inflows should be different. It is clear  that the ultimate reason
of the abnormal positions in the luminosity-metallicity relation should explain
the position  of all the four objects. We have to take into account that the
luminosity-metallicity relation used by \citet{R03} (as  well as the previous
one obtained by \citealt*{SKH89}) has been obtained for local dwarf irregular 
galaxies, with $M_B$ $\geq$ $-$18 (the most common upper limit for the
brightness of a dwarf galaxy). All the  galaxies with the ''unusual'' position
are brighter than this limit and therefore cannot strictly be considered dwarf 
galaxies, especially C, which is about 1.5 magnitudes brighter than that limit.
In fact, \citet{R03}  indicate that member C has a luminosity like that of a
typical late-type spiral galaxy, and also that its position in  the
luminosity-metallicity relation is compatible with the extreme low end of the
abundance range of late-type  spirals \citep{G97}. However, we think that the O
abundance of member C is more consistent with the expected abundance  in the
external zones of spiral galaxies of the same luminosity (see \citealt{G97}; 
their Figure 18).  We have derived the O abundance in the brightest
part of C, which should correspond to the nucleus of the  galaxy. Therefore,
the correct comparison would be with a relation between  luminosity and central
abundances, as it is shown in Figure 16 of \citet{G97}. In this case, member C,
as  well as the other three bright galaxies of the group are located between
0.5 and 1 dex below the central abundances  reported by \citet{G97}. In any
case, the O/H ratios of the brightest members of HCG~31 are more consistent 
with those found in Magellanic-type irregular galaxies. 

Several authors have questioned the validity of the luminosity-metallicity
relation of dwarf galaxies. \citet{CA93}  and \citet{PA93} find no evident
relation for blue compact galaxies, and \citet{M94} does  not support such
relationship for low surface brightness galaxies. Therefore, the
luminosity-metallicity relation may  be not valid for gas-rich galaxies. In
particular, HCG~31 should be considered as a gas-rich group because its global 
$L_B$/M(\ion{H}{1}) is approximately unity taking into account a total hydrogen
mass of 2.1 $\times$ 10$^{10}$  M$_\odot$  \citep{W91}. Chemical evolutionary
models by \citet*{HG03} indicate that a linear  correlation between luminosity
and metallicity could be feasible for non-bursting dwarf irregular galaxies, but
that a plateau is reached at about $M_B \leq$ $-$18, just at the limit of the
definition of a dwarf galaxy. This can be seen  graphically in the Figure 8 of
\citet{HG03}, where their most luminous model galaxies are located at 
12 + log (O/H) $\simeq 8.1$, precisely the abundance of the
brightest members of HCG~31. Other  interesting conclusion of \citet{HG03} is
that variations in the stellar mass-to-light ratio can  contribute
significantly to the scatter in the luminosity-metallicity relation. This was
first pointed out by  \citet{PA93}, who interpret the lack of
correlation between the $B$-luminosity and metallicity of  their sample of
\ion{H}{2} galaxies as the transient increase of luminosity due to the intense
starburst that these  objects  are experiencing. \citeauthor{PA93} propose that,
once the bursts evolve, the position of these galaxies will  move toward lower
luminosities, perhaps approaching the luminosity-metallicity relation of
non-bursting dwarf  irregulars. In fact, \citet{HGO98} propose that perhaps NIR
magnitude is a more suitable   transient indicator, taking into account the
contribution of the starburst to the $B$-magnitude. 

\begin{figure*}[h!]
\centering
\includegraphics[width=0.8\linewidth]{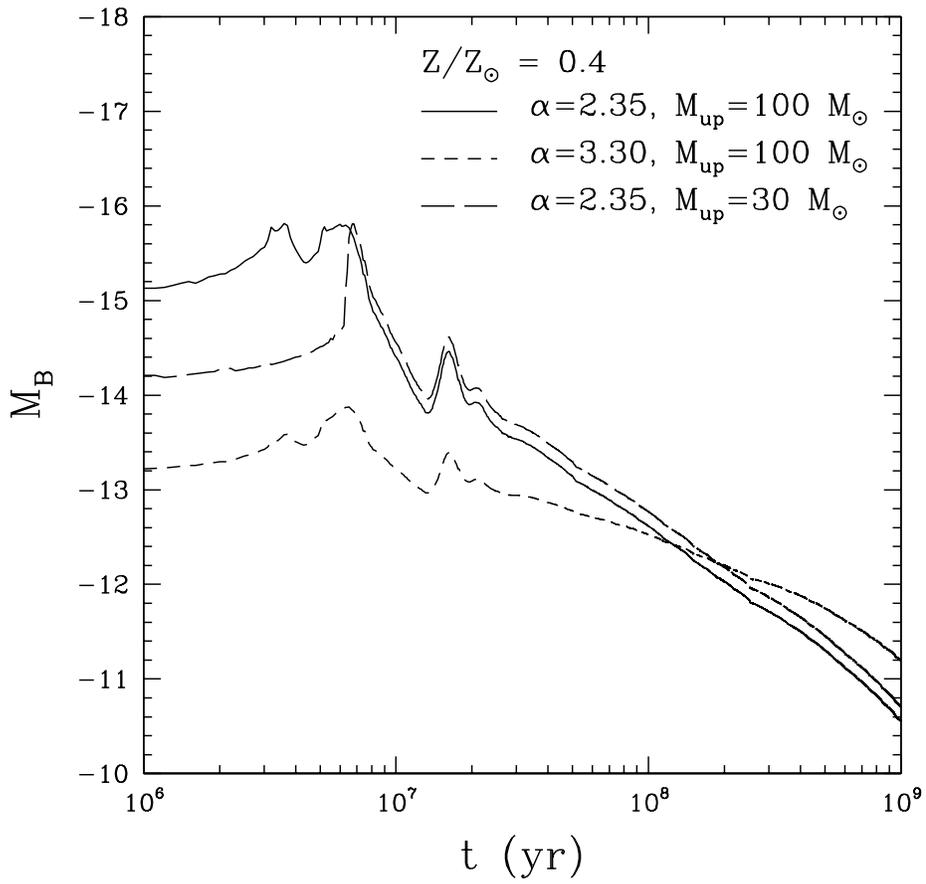}
\protect\caption[ ]{\small{$M_B$ evolution versus time for the indicated STARBURST~99 \citep{L99} models.}}
\label{fig15}
\end{figure*}

We have further explored the effect of the onset of the starbursts on the
absolute $B$-magnitude of a galaxy. In  Figure~\ref{fig15} we show the
evolution of $M_B$ versus time for a Z/Z$_\odot$ =0.2 burst as predicted by the
population  synthesis models of STARBURST~99 \citep{L99}. We have considered a
10$^6$ M$_\odot$ instantaneous burst  and three different stellar mass
distributions: a) Salpeter IMF and 100 M$_\odot$ upper mass; b) Salpeter IMF
and 30  M$_\odot$  upper mass; c) stepper IMF ($\alpha$ = 3.30) and 100
M$_\odot$ upper mass. We consider that case (a) could  be more appropriate for
the galaxies of HCG~31. In fact, as \citet*{EKT93} indicate, massive star
formation  in interacting galaxies may have a higher efficiency than in other
scenarios, and the clouds should have larger  internal temperatures, producing
the shift in the stellar mass function toward more massive stars. On the other
hand,  as \citet{J99} and \citet{JC00} demonstrated, the starbursts of NGC 1741
and other members of  HCG~31 are composed by a large number of SSCs. The masses
of those clusters range between 10$^4$ and 10$^6$  M$_\odot$, so the
$B$-magnitude evolution of the galaxies should be represented as a combination
of a number of  clusters evolving as shown in Figure~\ref{fig15}.
The different curves clearly indicate that the $B$-magnitude of the  burst 
could be increased in several magnitudes during the first 10 Myr with respect to
the brightness in the quiescent phase.  This effect is surely affecting the
positions of the galaxies of HCG~31 shown in Figure~\ref{fig14} because their
blue  luminosities are dominated by young, almost coeval,
starbursts. In the future, the positions of the  brightest galaxies (A, B, C,
and G) will be shifted toward the right on the diagram in their evolution
toward the  quiescent phase; eventually reaching positions nearer the
metallicity-luminosity relation. In this case, it is even  possible that some of
these galaxies could be classified as bona-fide dwarfs.


Summarizing, the members of HCG~31 are dominated by strong and very young
starbursts with ages below 10 Myr. We  consider that the emission of this
short-lived young stellar population is increasing so intensely their 
$B$-luminosity that the use of the metallicity-luminosity relation is no
longer appropriate for these objects. In  accord with other previous
suggestions, we propose that the use of this relation is not appropriate for 
starburst-dominated galaxies.

\subsection{Are E and F Tidal Dwarf Galaxies?}

In contrast with what happens for the brightest galaxies, the position of
objects E and F1 in the O/H versus $M_B$  diagram (Figure~\ref{fig14}) seems to
be rather consistent with the metallicity-luminosity relation. However, we
think  that  this could be fortuitous. Taking into account the conclusion of
the previous section, the very blue colors and small  sizes of those objects
imply that the contribution of the young population should be even more 
important than in the brightest galaxies of the group. Probably, the future
photometric evolution of their starbursts  will move their positions toward
lower luminosities, away from the relation, after the first 10 Myr. If this
prediction is  correct, it favors their tidal dwarf nature, as several authors
have pointed out previously. As it was commented  above, the O/H ratios of E,
F1, and F2 are of the order of the values obtained for the brightest members of
HCG~31  despite their different luminosities and distances to the
A+C complex. Moreover, E, F1, and F2 have also N/O  ratios (considering N/O
$\approx$ N$^+$/O$^+$) similar to those of the brightest galaxies within the
uncertainties  (see Table~\ref{table5}). This is  remarkable, because the bulk
of N is thought to be produced by intermediate mass stars, indicating that the
material  shows some degree of contamination of previous populations. This fact
favors that the galaxies are not made of  pristine clouds but instead of
material stripped from other chemically evolved galaxies. 

Tidal dwarf galaxies (TDGs) are thought to be built from the outer disks of
their parent galaxies. \citet*{WDA03}  present results for a large number of
TDG candidates in a sample of interacting galaxies. These authors  indicate
that their TDG candidates are likely to contain a significant amount of old
stars, and that the tidal tails  located at different distances of the same
parent galaxy do not show evidences of abundance gradients along the  tails. In
the case of members E and F, we do not see clear abundance differences within
the uncertainties, and the  presence of older population does not seem to be
important. The absence of differences in the O/H ratios between the  brightest
objects, specially the A+C complex, and members E and F, is remarkable in this
sense. If E and F are in  fact TDGs of one of the brightest galaxies of the
group (member A or C) and the progenitor was a spiral (with an  abundance
gradient), they should show lower O/H ratios than the nucleus of the parent
galaxy. This result favors that  the progenitor is probably an irregular galaxy
with an homogeneous abundance across its volume. 

\citet{WDA03} obtained a mean  12+log(O/H)=8.34$\pm$0.14 for their sample of
TDGs candidates. The O/H ratios of E, F1, and F2 (around 8.1) are compatible with the lower
end of that range of values, but are higher than the abundances of the  three
objects that \citet{WDA03} interpret as pre-existent dwarf companion galaxies
(12+log(O/H) between 7.6 and 7.8). A pre-existent dwarf galaxy should have an
oxygen abundance compatible with the metallicity-luminosity relation studied 
before. If members E, F1 and F2 in HCG~31 were pre-existent dwarf galaxies,
following this relation they would have metallicities around 0.5 dex lower than
the brightest galaxies of the group (between 7.6 and 7.7), but 
these values are not observed. On the other hand, \citet{WD01} and
\citet{WDA03} indicate that the three pre-existent companions seem  to contain
a strong old stellar component, property that is not observed in objects E, F1,
and F2 either.  

Normal dwarf galaxies are stable entities with their own dynamics, so the best
definition of a TDG is that it behaves  as a self-gravitating object
\citep{Du00, WD01}. To confirm if a knot in a tail is a genuine  TDG we have to
know the kinematics of the knot. \citet{WDA03} interpret a knot as a TDG
candidate when it  shows a kinematics decoupled from the expanding motion of
the tidal tail, and possibly rotation. In Figure~\ref{fig8},  we can  see that
the faint optical tidal tail that emerges from the southwest of member C shows
a linear continuous tidal  streaming motion just at the north of member E, with
velocities ranging from $-$100 to +50 km s$^{-1}$ from its base  to object H.
However, as it can be seen in Figure~\ref{fig8}b, member E shows a velocity
pattern very different from  the kinematical behavior of the tidal tail. This
fact indicates two possibilities: E does not belong to the faint  optical tidal
tail or is a TDG decoupled from the tail. The velocity patterns of objects F1
and F2 are also rather  constant and very different from those found in the
optical tidal tail, especially in the apparent tip, object H. However, there is
not  clear evidence of rotation patterns in the velocity
distributions inside objects F1 and F2. 

As we have commented in section 3.3.4, the mean velocities of objects F1 and F2
are similar to those of members E and  G, which coincide with the radial
velocity of the \ion{H}{1} cloud in this part of HCG~31. This fact suggests
that  perhaps  E, F1, and F2 are related to the arm-like \ion{H}{1} structure
that extends to the southeast of the A+C complex. We think  that the velocity
pattern and the morphology of the system are compatible with the presence of
two spatially  coincident kinematical structures: 
\begin{itemize}
\item the arm-like \ion{H}{1} structure that extends from A+C in direction to
member G, from which objects E and F may have formed, and showing a rather
constant radial velocity; 
\item the optical tidal tail that emerges from the southwest of the A+C complex, which consist of a curved string of 
faint star-forming regions that seems to end at the position of object H (see Fig.~\ref{fig1}). It shows a clear 
streaming motion bending away from us.
\end{itemize} 

We propose that objects E, F1, and F2 are TDGs made from material in
the southern arm-like \ion{H}{1}  extension, which was stripped from the parent
galaxy, probably the A+C complex because of the relatively high  chemical
abundances, very similar to those of the brightest galaxies of the group. The
apparent absence of old stellar population in the  three objects indicates that
they are basically made of gaseous (but chemically evolved) material. In fact,
a local  maximum in the distribution of \ion{H}{1} emission coincides with the
position of member F \citep{W91}. We have estimated that these dwarf objects 
have a total mass between 10$^7$ and
10$^8$ \Mo, the typical values for dwarf irregular galaxies. 
Therefore, objects E, F1, and F2 should have
formed from gas stripped from the disk of a parent galaxy with a rather  flat
abundance gradient, perhaps an irregular galaxy.

\subsection{The star formation history of HCG~31 revisited}

The history and future fate of the HCG~31 system have been discussed by several
authors \citep{R90, IPV97,  JC00, R03}. We consider that our new data give some
new useful elements that can lead to a better understanding of this amazing
interacting system. 

As it was commented in \S\S~3.1 and 3.3.4, the complex morphology of A+C
(double nucleus, presence of stripped disks  or tidal tails) and the
position-velocity diagram of PA 61$^\circ$ indicate that both galaxies are
merging.  In fact, this object shows the most intense (and one of the most
recent) star formation  activity of the group. Moreover, IV97 indicate that the
colors of the A+C complex are well reproduced by the models of  photometric
evolution of mergers of two galaxies of similar morphological type of
\citet{FG94}. \citet{O02} presented  maps in 7.7 and 14.3 $\mu$m and
spectrometry of HCG~31 using the Infrared Space  Observatory (\emph{ISO}). They
detected strong mid-IR emission from the central burst, along with strong
polycyclic  aromatic hydrocarbon (PAH) features and a blend of other features,
including [\ion{S}{4}] at 10.5 $\mu$m. The 14.3/6.75  $\mu$m and  14.3/7.7
$\mu$m flux ratios suggest that the central burst within HCG~31 may be moving
toward the post-starburst  phase. On the other hand, \citet{Y97} observed a CO
deficiency and a peculiar molecular gas distribution in HCG~31. The  brightest
CO peak occurs in the overlap region between the galaxies A and C, and the
enhanced CO emission and  associated star formation may be the result of the
ongoing merger. They interpreted the observations with a scenario  where the
compact group is subject to continuous tidal disruptions. Both tidally induced
star formation and tidal  stripping may be at work resulting in the reduced
molecular gas reserve. This is a very likely situation for HCG~31
taking into account the different and complex evidences of interaction,
merging, and tidal stripping that this system shows. \citet{Y97} also found a
similar behavior in the CO distribution of HCG 92 (\emph{Stephan's  Quintet}). 

Galaxy B shows solid-body rotation with an amplitude of about 200 km s$^{-1}$
(see the position-velocity diagram of  \citealt{R90}, that passes precisely
across the major axis of the galaxy), and is counter-rotating with  respect to
the A+C complex. Our broad-band images, as well as the H$\alpha$ ones of IV97,
indicate some physical  connection (perhaps a tidal bridge) between B and the
A+C complex (this connection is quite obvious in  Fig.~\ref{fig1}). \citet{R03}
describe B as a dwarf spiral or irregular galaxy seen nearly edge-on, but we
think  that it is a bar-like irregular galaxy, taking into account its 
morphology in H$\alpha$ (see IV97). It is clear from its nebular emission and
very blue colors that galaxy B is experiencing a strong star formation
activity. This activity may be related to the interaction with the A+C 
complex. IV97 found that this galaxy shows two maxima in H$\alpha$ equivalent
width located at the northern and  southern tips of the galaxy. They suggest
that this fact could be related to some kind of stripping mechanism. It is 
interesting to note that the behavior of galaxy B is quite similar to that of
the blue compact dwarf galaxy Mrk 1094  ($M_B$ = $-$18.4). This galaxy has been
studied by \citet{M99} and consists of an ensemble of star-forming  knots
distributed in a twisted bar-like structure, which also shows two maxima of
$W(H\alpha)$ at the tips of the  bar.  \citet{M99} propose that the
characteristics of Mrk 1094 can be explained if it is interacting with  a
companion \ion{H}{1} cloud located 50 kpc to the south. The models of tidal
interactions of disk galaxies of  \citet{NI86} predict that bars are very
likely produced in close encounters. These authors predict that the  maximum
activity of the bursts produced by the interactions takes place $\sim$ 3
$\times$ 10$^8$ yr after the  \emph{perigalacticon} (the time when cloud-cloud
collisions become more frequent and the SFR reaches a maximum),  which is of
the order of the crossing times for galaxies within groups the size of HCG~31
\citep{R90}.  According to \citet{N88}, the infall of gas toward the nuclear
region of the galaxy triggers the SF in the center  of the galaxy. Perhaps, the
maxima of $W(H\alpha)$ at the tips is indicating the propagation of the star
formation outwards along  the bar. As IV97 and \citet{R03} suggest, galaxy B
may probably merge with the A+C complex in a short time and  is now undergoing the
first stages of the interaction process. 

As discussed in \S~4.5, members E and F seem to be TDGs candidates
made of gas from the armlike  \ion{H}{1}  structure. This structure should have
originated in the external zones of the galaxies of the A+C complex, since
the chemical abundances in E, F, and the A+C complex are very similar. The
fact that this \ion{H}{1} structure and  the  optical tidal tail at the southwest of A+C 
are kinematically decoupled is striking, and indicates that they are different
objects.  \citet{JC00} suggest that both the optical tidal tail and the
arm-like structure are the same entity and could  correspond to the inflow of
gas to member G from the direction of the main body of the \ion{H}{1} cloud in
which  HCG~31 is  embedded, regardless of the direction of motion of G relative
to the ambient gas. This scenario can explain the  coincidence of member G with
a peak of the \ion{H}{1} distribution at the end of the arm-like structure.
\citet{R03}  propose that a gravitational interaction between member G and the
A+C complex is the most complete explanation  of the events occurring in
HCG~31. For these authors, the star formation processes taking place almost
simultaneously  in A+C, E, F, and G were produced by the fly-by encounter
between A+C and G. \citet{R03} interpret the \ion{H}{1}  structure as the
counterpart of the optical tidal tail, but we have found kinematical
evidence that this is  unlikely. In this context, the absence of a radial
velocity gradient along the arm-like \ion{H}{1} structure indicates  that  the
relative motion between G and the A+C complex must be in the plane of the sky.
\citet{R03} estimate that the  expected range of possible transverse velocities of
the gas is of several hundreds of km s$^{-1}$, and conclude that  neither the
velocities nor the spatial extension between galaxies G and A+C (40 kpc) are
unusual when compared to those found in other well-studied  interacting systems. 

The presence of two adjacent tails has been reported in other interacting
systems, but these reports correspond to bifurcated tails [e.g. M81 \citep*{vH79,
Y94}; NGC 3921 \citep{H96}; NGC 2535/6 \citep{K97}; Arp 299  \citep{H99}; NGC
4038/39 \citep{H01}]. For example,  in the case of NGC 4038/39 (\emph{The
Antennae}) one of the tails is visible in the optical while the other, that 
runs parallel to the first, emits only in \ion{H}{1} and does not show an
optical counterpart. \citet{H01} interpret  the bifurcated tails of NGC 4038/39
as produced by a lateral twist in the tail that may cause the  outer edge
(gas-rich region) to lie in a different plane from the brighter and gas-poor
inner regions. This effect  can be exacerbated by a preexisting warp in the
progenitor disk. In all the cases with enough  kinematical data,
bifurcated tails show quite similar kinematics, in contrast to the situation we
are seeing  in HCG~31, where the southwest optical tail and the arm-like
\ion{H}{1} structure have different kinematics.    

Considering all the observational data available for the group, we propose that
there are several simultaneous  interaction processes taking place in HCG~31.
The fly-by encounter between G and the A+C complex originated the  \ion{H}{1} 
tidal tail (arm-like structure). This process perhaps occurred before the
merging of A and C because the \ion{H}{1}  tail is  the longest structure in
the group. Afterwards, the merging of A and C would create the northeast and
southwest  optical tidal tails. The triggering of the star formation in galaxy
G could be due to the inflow of the material from  the \ion{H}{1} tidal tail.
In fact, the star-forming regions form a U shape along the northwest side of G,
precisely  in the  direction toward the A+C complex. The TDGs candidates E and
F have been recently created from interaction-induced instabilities on the
\ion{H}{1} tidal tail and nowadays they show a strong star formation activity,
specially intense in F where WR features are detected. 
Finally, galaxy B suffers some kind of
interaction-induced star formation  probably propagating along its bar. 

As we can see, HCG~31 is a very complex system where interaction effects are
not only between two galaxies. In fact, four galaxies (A, C, B, and G) seem
to be involved in different interaction processes. This does not seem to be a 
common phenomenon in the local universe, but it could be very usual in places
with a high density of galaxies, like  the early universe, when interactions
could occur frequently. Nowadays, hierarchical galaxies formation models (e.g. 
\citealt{KW93}) suggest this scenario. The detailed study of systems like HCG~31
in the local universe could  provide important clues about the evolution of
galaxies in high density environments.

\section{Conclusions}

In this paper we present results based on new broad-band optical and
near-infrared imagery and optical spectroscopy  of the compact group HCG~31.
Our main results can be summarized as follows:
\begin{itemize}
\item We have obtained remarkably consistent age estimations for the most
recent star formation bursts of the  group using different observational
indicators and theoretical models. Member F hosts the youngest starburst of the
group (age between 2 and 3 Myr). It does not show evidences of an underlying
old stellar population and hosts a substantial  population of WR stars. 
\item We have obtained direct determinations of the oxygen abundance for most of
the galaxies of the group,  finding very similar values in all the cases
(12+log(O/H) = 8.0--8.2) despite their very different  absolute magnitudes. We
report the first direct abundance determination for members B, E, F2, and G.
\item We detect the \ion{C}{2} $\lambda$4267 line in the spectrum of member C.
This is the first time this  recombination line is reported in an \ion{H}{2}
galaxy. The C$^{++}$/H$^+$ ratio we derive from this line is  too
high, but this could be due to our possible overestimation of the intensity of
this weak line.
\item We argue that the use of traditional metallicity-luminosity relations
based on the absolute  $B$-magnitude (as those of \citealt{RM95}, constructed for
non-bursting dwarf galaxies) is not appropriate for  dwarf starburst galaxies.
Their luminosity is dominated by the transient contribution of the starburst
to the blue  luminosity, which increases dramatically the $B$-magnitude with
respect to the pre- or post-starburst quiescent phase.
\item We propose that members E and F are tidal dwarf galaxies made of
gas-rich material stripped  from the A+C complex because of: (a) their high
oxygen abundance (similar to that of the other galaxies of the group);  (b) the
absence of substantial underlying old stellar populations; (c) their
kinematics. 
\item We propose that HCG~31 is a very complex interacting system of galaxies
that is suffering several  simultaneous interaction processes. A and C are the
dominant members of the group and they are now merging.  This merging process
could be the origin of the two optical tidal tails that extend towards the
northeast and  southwest of the A+C complex. There was a fly-by encounter
between G and the A+C complex (or one of these galaxies  before the merging) and
this encounter produced an \ion{H}{1} tidal tail from the stripping of the
external gas of  A+C.  This tail moves in or very close to the direction of the
plane of the sky.   Members E and F could have formed from instabilities produced
in this gas-rich structure. Finally, member B is in  interaction with the A+C
complex. This is indicated by: (a) the strong star formation that is taking
place along the   tilted disk of B; (b) the presence of apparent streaming
motions in the ionized gas between both  objects; (c) the clear faint bridge of
matter that connects B and the A+C complex. 
\end{itemize}

\acknowledgments

We would like to thank Jos\'e A. Caballero for providing us with a macro
for NIR reduction. We thank Ver\'onica Melo and Ismael Mart\'{\i}nez-Delgado
for permitting us to observe at CST on 2003 February 4. We are very grateful to 
Jorge Garc\'{\i}a-Rojas for his help in the derivation of the helium
abundances. We would also like to  thank David
Mart\'{\i}nez-Delgado for permitting us to observe at INT on 2003 September 22 
and, especially, the anonymous referee for his/her valuable comments.
M.R. acknowledges support from Mexican CONACYT project J37680-E.

\clearpage 

\end{document}